%% file: arXiv_v3.tex
\documentclass[a4paper,11pt]{article}
\pdfoutput=1 

\usepackage{jheppub} 
\makeatletter
\DeclareRobustCommand*{\bfseries}{%
  \not@math@alphabet\bfseries\mathbf
  \fontseries\bfdefault\selectfont
  \boldmath
}
\makeatother

\usepackage{calc}
\usepackage{rotating}
\usepackage[english]{babel}
\usepackage[utf8]{inputenc}
\usepackage{graphicx}
\usepackage{subfig}
\usepackage{float}
\usepackage{amsmath}
\usepackage{amssymb}
\usepackage{amsthm}
\usepackage{latexsym}
\usepackage{dcolumn}
\usepackage{hyperref}
\input macros.tex

\newcolumntype{P}[1]{>{\centering\arraybackslash}p{#1}}

\restylefloat{figure}

\title{Entanglement in flavored scalar scattering}

\author{Kamila Kowalska}
\author{and Enrico Maria Sessolo}

\affiliation{National Centre for Nuclear Research,\\
Pasteura 7, 02-093 Warsaw, Poland}

\emailAdd{kamila.kowalska@ncbj.gov.pl}
\emailAdd{enrico.sessolo@ncbj.gov.pl}

\abstract{We investigate quantum entanglement in high-energy $2\to 2$ scalar scattering, where the scalars are characterized by an internal flavor quantum number acting like a qubit. Working at the 1-loop order in perturbation theory, we build the final-state density matrix as a function of the scattering amplitudes connecting the initial to the outgoing state. In this construction, the unitarity of the $S$-matrix is guaranteed at the required order by the optical theorem. We consider the post-scattering entanglement between the momentum and flavor degrees of freedom of the final-state particles, as well as the entanglement of the two-qubit flavor subsystem. In each case we identify the couplings of the scalar potential that can generate, destroy, or transfer entanglement between different bipartite subspaces of the Hilbert space. }

\begin{document}
\maketitle

\setcounter{footnote}{0}

\section{Introduction}

Quantum entanglement is the fundamental property of nature allowing one to set apart classical and quantum phenomena. A composite quantum system is called entangled if it cannot be written as a tensor product of its individual components, which results in the violation of Bell inequalities\cite{PhysicsPhysiqueFizika.1.195,PhysRevLett.47.460}.

In quantum field theory~(QFT), composite systems of particular interest are those describing interacting particles. All properties of a scattering event are captured in this framework by the $S$-matrix, which acts as a unitary operator in the Hilbert space spanned by the state vectors associated with scattered particles. A unitary transformation from the initial to the final (scattered) state may create (or destroy) the entanglement in the system and the question arises how variations in entanglement are related to the structure and properties of the $S$-matrix.

In perturbation theory, the issue was addressed in Refs.\cite{Balasubramanian:2011wt,Hsu:2012gk,Seki:2014cgq,Peschanski:2016hgk,Carney:2016tcs,Peschanski:2019yah} for generic two-particle scattering. Initial and final states were represented by vectors in the infinite-dimensional Hilbert space of momenta and the reduced density matrix  was derived at the lowest order in the weak coupling. It was found that the scattered system would always feature entanglement between various modes of momenta, as long as the interaction is turned on. More quantitatively, the entanglement entropy of the composite system in the momentum space is proportional to the total scattering cross section. Higher-order effects were studied in Ref.\cite{Faleiro:2016lsf}.

In more realistic settings, scattered particles will usually carry internal quantum numbers like spin, flavor, polarization, etc. The  Hilbert space of momentum should thus be extended by extra finite-dimensional factors, which may feature their own entanglement or add to the overall  entanglement of the scattered system. Refs.\cite{He:2007du,Ratzel:2016qhg,Fan:2017hcd,Fan:2017mth,Araujo:2019mni,Fonseca:2021uhd,Fan:2021qfd,Shivashankara:2023koj,Blasone:2024dud,Blasone:2024jzv} studied the resulting entanglement entropy in various QFTs.  Alternatively, momentum can be entirely removed from the picture by projecting onto specific momenta of the final state. Along these lines, Refs.\cite{Cervera-Lierta:2017tdt,Fedida:2022izl} analyzed the helicity entanglement in quantum electrodynamics (QED).

One may also wonder whether the ability of a given $S$-matrix to create or destroy entanglement in a scattering event might point towards a particular symmetry of the QFT under consideration. While this enticing question was answered positively in the framework of nuclear theory and low-scale quantum chromodynamics\cite{Beane:2018oxh,Beane:2020wjl,Low:2021ufv,Beane:2021zvo,Liu:2022grf}, in the realm of high-energy particle physics we are still far from definite conclusions. Reference\cite{Carena:2023vjc} analyzed the issue of entanglement in flavor space in the framework of the two-Higgs-doublet model (2HDM), highlighting a connection between minimal entanglement of the scalar scattering final state and the global SO(8) symmetry of the scalar potential.\footnote{As we shall see, this conclusion may not hold if some of the assumptions made in Ref.\cite{Carena:2023vjc} were lifted.} In Ref.\cite{Cervera-Lierta:2017tdt} it was demonstrated that the requirement of maximal entanglement potentially allows for the emergence of gauge invariance in the vertex structure of QED. 
Finally, in a more general spirit, Refs.\cite{Cheung:2023hkq,Aoude:2024xpx} related the change in entanglement entropy with the positivity of the scattering amplitude.

Interestingly, in recent years progress was also made in our understanding of how to detect and test entanglement at colliders. This issue was analyzed for top-pair production\cite{Afik:2020onf,Fabbrichesi:2021npl,Severi:2021cnj,Afik:2022kwm,Aoude:2022imd,Aguilar-Saavedra:2022uye,Fabbrichesi:2022ovb,Severi:2022qjy,Dong:2023xiw,Varma:2023gwh,Aguilar-Saavedra:2023hss,Han:2023fci,ATLAS:2023fsd,Maltoni:2024tul,Aguilar-Saavedra:2024hwd,Duch:2024pwm}, for Higgs bosons decays\cite{Barr:2021zcp,Altakach:2022ywa,Ma:2023yvd,Ehataht:2023zzt}, and in the systems of massive gauge bosons\cite{Barr:2021zcp,Barr:2022wyq,Aguilar-Saavedra:2022wam,Ashby-Pickering:2022umy,Fabbrichesi:2023cev,Aoude:2023hxv,Bi:2023uop,Bernal:2023ruk} (see also Ref.\cite{Barr:2024djo} for a recent review).

In this study, we revisit the issue of the 
possible fundamental connection between the 
entangling properties of the $S$-matrix and the characteristics of the Lagrangian of some physics beyond the Standard Model~(BSM). 
As a specific example, we focus on the high-energy scattering of two flavored scalar particles in the 2HDM, for which we adopt the flavor internal quantum number as a qubit. Our study thus extends the findings of Ref.\cite{Carena:2023vjc}, with respect to which we employ a different approach that relies on the unitarity of the $S$-matrix in the Hilbert space of both flavor and momentum at the 1-loop order in perturbation theory. In this regard, it is important to point out that different measures of entanglement have to be used depending on whether the final flavor state is pure or a statistical admixture and that depends on the treatment of the momentum degrees of freedom. Note that in adopting this procedure, our results are also complementary to the findings of Ref.\cite{Aoude:2024xpx}.

It should be mentioned that the question of \textit{which} entanglement one is interested in quantifying is not trivial if the quantum state is characterized by internal quantum numbers as well as momentum. A bipartition of the entire multi-factor Hilbert space can be performed in this case across different subsystems, with each partition carrying its own information about the quantum nature of the final state. In one choice, the momentum space could be entirely traced out and the resulting entanglement entropy will indicate whether the scattering final state can be written as a tensor product of the momentum and flavor Hilbert spaces. The intrinsic entanglement between flavor degrees of freedom in the remaining (untraced) 
flavor space can be analyzed subsequently, with the hope of extracting information on the interaction structure of the BSM Lagrangian. 

Alternatively, one may choose to trace out all the degrees of freedom of one of the scattered particles. While the entanglement of the two final states can be quantified in this way (see, \textit{e.g.}, Ref.\cite{Aoude:2024xpx} for the corresponding derivation), the result will not give much insight on the interaction structure of the BSM model, 
due to the intrinsic entanglement 
between momentum modes and internal degrees of freedom. 

Finally, we can project the scattering final state onto a particular  choice of momentum state (a ``measurement'' so to say), in the spirit of Refs.\cite{Cervera-Lierta:2017tdt,Fedida:2022izl}, where this was done for the helicity states of the electron in QED.
In this approach, the maximal and/or minimal entanglement conditions can be easily formulated in terms of the couplings of the BSM Lagrangian, although one should be careful when deriving generic conclusions regarding the model as any measurement affects the properties of a quantum state.

The paper is organized as follows. In \refsec{sec:scat} we derive the density matrix for a generic perturbative $2\to 2$ scattering process, in which the scattered quantum states carry internal discrete quantum numbers. We further examine various approaches to quantifying the entanglement in the final state. Special emphasis is given to the case of entanglement between momentum and internal degrees of freedom, and the case of entanglement between subspaces 
reproducing a traditional qubit structure.
In \refsec{sec:2hdm} we apply this formalism to the 2HDM for which we identify -- in similar fashion to Ref.\cite{Carena:2023vjc} -- the internal quantum number with the flavor of the electroweak scalar doublets. We study $2\to 2$ scattering processes of this model at high energy. At the leading order, they proceed through contact interactions, allowing us to distinguish the specific quartic couplings that create entanglement or modify its distribution in the system. We summarize our findings in \refsec{sec:sum}. Two appendices feature, respectively, a discussion of the physical meaning of the momentum-space normalization factor and the derivation of the density matrix reduced over one of the subsystems.

\section{Quantum properties of $\boldsymbol{2\to 2}$ scattering}\label{sec:scat}

\subsection{Scattering amplitude and the density matrix\label{sec:sca_den}}

We consider in this work the scattering of two flavored particles at high energy. Our Hilbert space is 
\be
\mathcal{H}_{\textrm{tot}}=L^2(\mathbb{R}^3\otimes \mathbb{R}^3)\otimes \mathbb{C}^4\,,
\ee
composed of the \textit{momentum} Hilbert space of the scattering particles,  $L^2(\mathbb{R}^3\otimes \mathbb{R}^3)= L^2(\mathbb{R}^3)\otimes L^2(\mathbb{R}^3)$, and of a discrete-index \textit{flavor} Hilbert space built out of two qubits, $\mathbb{C}^4=\mathbb{C}^2\otimes \mathbb{C}^2$. 
A generic basis for $\mathcal{H}_{\textrm{tot}}$ will be indicated as $|\mathbf{p}_1 \mathbf{p}_2 \rangle |\alpha\beta\rangle$, where $\mathbf{p}_1, \mathbf{p}_2$ are continuous 3-dimensional variables and $\alpha,\beta$ can take two values, let us say~1 and~2. 

We normalize the momentum-basis vector as
\be\label{eq:norconv}
\langle\mathbf{p}_1\mathbf{p}_2 | \mathbf{p}_1' \mathbf{p}_2' \rangle
= (2\pi)^6\,4 E_{1} E_{2}\, 
\delta^3(\mathbf{p}_1-\mathbf{p}_1')\, \delta^3(\mathbf{p}_2-\mathbf{p}_2')\,,
\ee
where $E_1$, $E_2$ are the 0-th components of momentum 4-vectors $p_{1,2}=(E_{1,2},\mathbf{p}_{1,2})$ in Minkowski space. Equal indices in \refeq{eq:norconv} define the infinite normalization volume $V$, which in the center-of-mass frame is given by
\be
V=\langle\mathbf{p}_1\mathbf{p}_2 | \mathbf{p}_1 \mathbf{p}_2 \rangle = s\left[(2\pi)^3\,\delta^3(0)\right]^2\,,
\ee
where $s=4 E_1 E_2$ is the usual Mandelstam variable. 
The discrete, computational-basis vector is normalized trivially as
\be\label{eq:normfl}
\langle\alpha\beta | \gamma\delta\rangle=\delta_{\alpha\gamma}\,\delta_{\beta\delta}\,.
\ee

We consider a generic initial two-particle state before scattering, 
\bea\label{eq:instate}
|\textrm{in}\rangle &=&\left( \prod_{i=1,2}\int\frac{d^3 p_i}{(2\pi)^3}\frac{1}{\sqrt{2 E_i}}\right)
\phi_{A}(\mathbf{p}_1)\phi_B(\mathbf{p}_2)\,|\mathbf{p}_1\mathbf{p}_2\rangle\sum_{\alpha,\beta=1,2}a_{\alpha\beta}|\alpha\beta\rangle\nonumber\\
 &\approx & \frac{1}{\sqrt{V}}\,\sum_{\alpha,\beta=1,2}a_{\alpha\beta}|\mathbf{p}_A\mathbf{p}_B\rangle |\alpha\beta\rangle\,,
\eea
separated in momentum, but not necessarily separated in the discrete-index space $\mathbb{C}^4$. As a consequence of \refeq{eq:normfl},
$\sum_{\alpha,\beta}|a_{\alpha\beta}|^2=1$. The last line of \refeq{eq:instate} formalizes our working assumption that the wave packets $\phi_A(\mathbf{p}_1)$ and $\phi_B(\mathbf{p}_2)$ are very well peaked around well-defined initial momenta $\mathbf{p}_A$ and $\mathbf{p}_B$, respectively, so that they can  be formally expressed as
\be\label{eq:wavep}
\phi_{A,B}(\mathbf{p})=\sqrt{\frac{(2\pi)^3}{\delta^3(0)}}\,
\delta^3(\mathbf{p}-\mathbf{p}_{A,B})\,.\footnote{Strictly speaking, the $|\textrm{in}\rangle$ state thus defined does not belong to the momentum Hilbert space 
$L^2(\mathbb{R}^3\otimes \mathbb{R}^3)$, as it is a tempered distribution acting on a dense subspace of $L^2(\mathbb{R}^3\otimes \mathbb{R}^3)$. Equation~(\ref{eq:wavep}) should therefore be interpreted as an approximation for the physical wave packet.}
\ee

The elements of the $S$-matrix in the basis $\{\langle \mathbf{p}_i \mathbf{p}_j|\langle \gamma\delta |$, $|\mathbf{p}_a \mathbf{p}_b \rangle |\alpha\beta\rangle\}$ are given by
\begin{multline}\label{eq:Smat}
S_{\gamma\delta\alpha\beta}^{ijab}=\left(\mathcal{I}+iT\right)_{\gamma\delta\alpha\beta}^{ijab}\\
=(2\pi)^6\,4\,E_a E_b\, \delta_{\gamma\delta\alpha\beta}^{ijab}
+(2\pi)^4\delta^4(p_a+p_b-p_i-p_j)\,
i\mathcal{M}_{\gamma\delta,\alpha\beta}(p_a,p_b\to p_i,p_j)\,,
\end{multline}
where 
\be
\delta_{\gamma\delta\alpha\beta}^{ijab}=\delta_{\alpha\gamma}\,\delta_{\beta\delta}\,\delta^3(\mathbf{p}_a-\mathbf{p}_i)\,\delta^3(\mathbf{p}_b-\mathbf{p}_j)\,,
\ee
and the scattering amplitude $\mathcal{M}$ carries discrete indices 
$\alpha\beta\gamma\delta$ besides momentum indices. 

Starting with the generic initial state of \refeq{eq:instate}, the post-scattering final state reads
\be\label{eq:outst}
|\textrm{out}\rangle = S|\textrm{in}\rangle=\sum_{\gamma\delta}\int\int
\frac{d^3 p_i}{(2\pi)^3}\frac{1}{2 E_i}\frac{d^3 p_j}{(2\pi)^3}\frac{1}{2 E_j}\, \psi_{\gamma\delta}(\mathbf{p}_i,\mathbf{p}_j)
|\mathbf{p}_i \mathbf{p}_j \rangle |\gamma\delta\rangle\,,
\ee
where we have introduced the final-state wave functions,
\begin{multline}\label{eq:wave_fun}
\psi_{\gamma\delta}(\mathbf{p}_i,\mathbf{p}_j)= \sum_{\alpha\beta} \int\int
\frac{d^3 p_a}{(2\pi)^3 \sqrt{2 E_a}}\,\frac{d^3 p_b}{(2\pi)^3 \sqrt{2 E_b}}\,   S_{\gamma\delta\alpha\beta}^{ijab}\, a_{\alpha\beta}\, \phi_A(\mathbf{p}_a) \phi_B(\mathbf{p}_b)\\
=a_{\gamma\delta}\, \sqrt{4\,E_i E_j}\, \phi_A(\mathbf{p}_i)\phi_B(\mathbf{p}_j) \\
+\int\int
\frac{d^3 p_a}{(2\pi)^3}\,\frac{d^3 p_b}{(2\pi)^3}\, \frac{\phi_A(\mathbf{p}_a)}{\sqrt{2 E_a}}\frac{\phi_B(\mathbf{p}_b)}{\sqrt{2 E_b}}  
(2\pi)^4\delta^4(p_a+p_b-p_i-p_j)\, \sum_{\alpha\beta} 
i\mathcal{M}_{\gamma\delta,\alpha\beta}(p_a,p_b\to p_i,p_j)\, a_{\alpha\beta}\,.
\end{multline}
In the specific case of a very well-peaked initial-state wave packet~(\ref{eq:wavep}),
one finds
\begin{multline}\label{eq:spec}
\psi_{\gamma\delta}(\mathbf{p}_i,\mathbf{p}_j)
=a_{\gamma\delta} \sqrt{4\,E_i E_j}\,(2\pi)^3\,  \frac{\delta^3(\mathbf{p}_i-\mathbf{p}_A) \delta^3(\mathbf{p}_j-\mathbf{p}_B) }{\delta^3(0)}\\
+\frac{(2\pi)^4\delta^4(p_A+p_B-p_i-p_j)}{\sqrt{V}}i\sum_{\alpha\beta} \mathcal{M}_{\gamma\delta,\alpha\beta}(p_A,p_B\to p_i,p_j)\, a_{\alpha\beta}\,.
\end{multline}

We point out that the $|\textrm{out}\rangle$ state is properly normalized order-by-order 
in perturbation theory, as a consequence of the optical theorem. Recalling Eqs.~(\ref{eq:outst}) and 
(\ref{eq:spec}) one writes
\bea\label{eq:norm}
\langle\textrm{out}|\textrm{out}\rangle &=& 
\sum_{\gamma\delta}\int\int \frac{d^3 p_i}{(2\pi)^3}\frac{1}{2 E_i}\frac{d^3 p_j}{(2\pi)^3}\frac{1}{2 E_j}\, |\psi_{\gamma\delta}(\mathbf{p}_i,\mathbf{p}_j) |^2 \nonumber\\
 &=&1 +\Delta\left(i\sum_{\alpha\beta,\gamma\delta}a^{\ast}_{\alpha\beta} \mathcal{M}_{\alpha\beta,\gamma\delta}(p_A,p_B\to p_A, p_B) a_{\gamma\delta}+\textrm{c.c.}\right) \nonumber\\
 & &+\,\,\Delta \int\int  \frac{d^3 p_i}{(2\pi)^3}\frac{1}{2 E_i}
 \frac{d^3 p_j}{(2\pi)^3}\frac{1}{2 E_j}
 (2\pi)^4 \delta^4(p_A+p_B-p_i-p_j) \nonumber \\
 & &\times \sum_{\alpha\beta,\rho\epsilon,\sigma\tau}
\mathcal{M}_{\alpha\beta,\rho\epsilon}(p_A,p_B\to p_i,p_j) a_{\rho\epsilon}
 \mathcal{M}^{\ast}_{\alpha\beta,\sigma\tau}( p_A,p_B \to p_i,p_j)  a^{\ast}_{\sigma\tau}\,,
\eea
where we have introduced an indeterminate normalization factor, 
\be
\Delta=\frac{(2\pi)^4\delta^4(p_A+p_B-p_A-p_B)}{4 E_A E_B\left[(2\pi)^3\,\delta^3(0)\right]^2}=\frac{\delta(0)}{s\,(2\pi)^2\,\delta^3(0)}\,.
\ee
In \refeq{eq:norm}, the imaginary part of the forward amplitude (in parentheses) 
cancels out, order by order, with the last addend after integrating the latter over phase space.

In the basis $\{\langle\mathbf{p}_a \mathbf{p}_b  |\langle \alpha\beta|,
| \mathbf{p}_i \mathbf{p}_j \rangle| \gamma\delta \rangle\}$, 
the entries of the density matrix $|\textrm{out}\rangle\langle \textrm{out}|$, corresponding to the final state after scattering,
can be calculated from the wave function straightforwardly:
\be\label{eq:rho}
\rho_{\alpha\beta,\gamma\delta}(\mathbf{p}_a,\mathbf{p}_b;\mathbf{p}_i,\mathbf{p}_j)=
\psi_{\alpha\beta}(\mathbf{p}_a,\mathbf{p}_b)\psi^{\ast}_{\gamma\delta}(\mathbf{p}_i,\mathbf{p}_j)\,.
\ee
The total trace of the density matrix,
\be
\textrm{Tr}(|\textrm{out}\rangle\langle \textrm{out}|)=\sum_{\alpha\beta}\int\int \frac{d^3 p_a}{(2\pi)^3}\frac{1}{2 E_a}\frac{d^3 p_b}{(2\pi)^3}\frac{1}{2 E_b}\,\rho_{\alpha\beta,\alpha\beta}(\mathbf{p}_a,\mathbf{p}_b;\mathbf{p}_a,\mathbf{p}_b)\,,
\ee
is 1 by means of the 
optical theorem, in exact equivalence to \refeq{eq:norm}. Similarly, it is also straightforward to see that the $|\textrm{out}\rangle$ state is a \textit{pure} quantum state, 
as $\textrm{Tr}\,(|\textrm{out}\rangle\langle \textrm{out}|)^2=1$. 

\subsection{Tracing out momentum degrees of freedom}\label{sec:traced}

Since in this work we are ultimately interested 
in computing the entanglement of quantum states in the two-qubit flavor space after the scattering process, we proceed to tracing out the momentum degrees of freedom from the density matrix
$|\textrm{out}\rangle \langle\textrm{out}|$. One should note, however, that the full Hilbert space $\mathcal{H}_{\textrm{tot}}$ 
could be decomposed in any chosen way
into a tensor product of two subspaces, each bipartition giving potentially rise to entanglement.

In this subsection, we 
trace over the entire momentum space, so to 
obtain the reduced density matrix for the discrete-index space $\mathcal{H}_{\textrm{flav}}\equiv\mathbb{C}^4$. Referring to the density matrix~(\ref{eq:rho}), one finds, in the $\{ \langle \alpha\beta|,
| \gamma\delta \rangle\}$ basis,
\begin{multline}\label{eq:rhoredF}
\tilde{\rho}_{\alpha\beta,\gamma\delta}=\int\int \frac{d^3 p_i}{(2\pi)^3}\frac{1}{2 E_i}\frac{d^3 p_j}{(2\pi)^3}\frac{1}{2 E_j}\,\rho_{\alpha\beta,\gamma\delta}(\mathbf{p}_i,\mathbf{p}_j;\mathbf{p}_i,\mathbf{p}_j)
=a_{\alpha\beta}\,a^{\ast}_{\gamma\delta}\\
+\Delta \left[-i\,  a_{\alpha\beta} \sum_{\epsilon\rho} \mathcal{M}^{\ast}_{\gamma\delta,\epsilon\rho}(p_A,p_B\to p_A, p_B) \,a^{\ast}_{\epsilon\rho}
+i\,a^{\ast}_{\gamma\delta}  \sum_{\epsilon\rho} \mathcal{M}_{\alpha\beta,\epsilon\rho}(p_A,p_B\to p_A, p_B) \,a_{\epsilon\rho}\right]\\
+\Delta \int \int \frac{d^3 p_i}{(2\pi)^3}\frac{1}{2 E_i}\frac{d^3 p_j}{(2\pi)^3}\frac{1}{2 E_j} 
 (2\pi)^4 \delta^4(p_A+p_B-p_i-p_j)  \\
\times \sum_{\epsilon\rho,\tau\sigma} 
\mathcal{M}_{\alpha\beta,\epsilon\rho}(p_A,p_B\to p_i,p_j)
\mathcal{M}^{\ast}_{\gamma\delta,\tau\sigma}(p_A,p_B \to p_i,p_j) \,a_{\epsilon\rho} \,a^{\ast}_{\tau\sigma}\,.
\end{multline}

As we showed at the end of \refsec{sec:sca_den}, $|\textrm{out}\rangle$ is a pure quantum state. We can then quantify right away the degree of entanglement between its momentum and flavor spaces 
by calculating the von~Neuman entropy of the reduced density matrix $\tilde{\rho}$\,:
\be\label{eq:vne}
S_N(\tilde{\rho})=-\sum_{i=1}^4\theta_i\log_2\theta_i\,,
\ee
where $\theta_i$ are the eigenvalues of $\tilde{\rho}$.
Incidentally, non-zero entanglement between the scattered particles signalizes that the reduced final state is not pure. Indeed one finds, up to the $\mathcal{M}^2$ order,
\begin{multline}\label{eq:rhosq}
\textrm{Tr}(\tilde{\rho}^2)=\sum_{\alpha\beta,\gamma\delta}\tilde{\rho}_{\alpha\beta,\gamma\delta}\,\tilde{\rho}_{\gamma\delta,\alpha\beta}=1+2\,\Delta \left[i\sum_{\alpha\beta,\epsilon\rho} \mathcal{M}_{\alpha\beta,\epsilon\rho}(p_A,p_B\to p_A, p_B) \,a^{\ast}_{\alpha\beta}a_{\epsilon\rho}+\textrm{c.c.}\right.\\
+\left. \int \int \frac{d^3 p_i}{(2\pi)^3}\frac{1}{2 E_i}\frac{d^3 p_j}{(2\pi)^3}\frac{1}{2 E_j} 
 (2\pi)^4 \delta^4(p_A+p_B-p_i-p_j)\right.\\
\left. \times\sum_{\epsilon\rho,\gamma\delta,\tau\sigma,\alpha\beta} 
\mathcal{M}_{\alpha\beta,\epsilon\rho}(p_A,p_B\to p_i,p_j)
\mathcal{M}^{\ast}_{\gamma\delta,\tau\sigma}(p_A,p_B \to p_i,p_j) \,a_{\gamma\delta}\,a^\ast_{\alpha\beta}\,a_{\epsilon\rho} \,a^{\ast}_{\tau\sigma}\right]\\
-\Delta^2\,\left[\sum_{\alpha\beta,\gamma\delta,\epsilon\rho,\tau\sigma} \mathcal{M}_{\alpha\beta,\epsilon\rho}(p_A,p_B\to p_A, p_B)\mathcal{M}_{\gamma\delta,\tau\sigma}(p_A,p_B\to p_A, p_B) \,a^{\ast}_{\alpha\beta}\,a_{\epsilon\rho}\,a^{\ast}_{\gamma\delta}\,a_{\tau\sigma}+\textrm{c.c.}\right.\\
-\left. 2\,\sum_{\alpha\beta,\epsilon\rho,\tau\sigma}\mathcal{M}_{\alpha\beta,\epsilon\rho}(p_A,p_B\to p_A, p_B)\,\mathcal{M}^\ast_{\alpha\beta,\tau\sigma}(p_A,p_B\to p_A, p_B)a^{\ast}_{\tau\sigma}  a_{\epsilon\rho}\right]\,.
\end{multline}

Note that the positive definiteness of the density matrix requires $\Delta\leq 1/(16\pi)$, as the last line in \refeq{eq:rhosq} is positive and may break unitarity
if $\Delta$ is too large. 
We shall consider in \refsec{sec:encrea}
a specific example of \refeq{eq:rhosq} with contact interactions, where the reader will be able to check this statement directly.
Because of this reason, our perturbative results will be expressed in a truncated power series of $\Delta$, besides in powers of the coupling constant. We discuss in some detail the physical meaning of $\Delta$ in Appendix~\ref{app:delta}. Finally, one 
can show that, when $\Delta< 1/(16\pi)$, 
$\tilde{\rho}$ represents a mixed quantum state of two qubits in the discrete flavor space.
Therefore, in this case, in order to quantify the entanglement in $\mathcal{H}_{\textrm{flav}}$ space after the scattering event, measures alternative to the von~Neumann entropy ought to be employed. Incidentally, if the scattered particles do not carry any additional quantum number apart from momentum (see, \textit{e.g.}, Ref.\cite{Seki:2014cgq}), the terms proportional to $\Delta^2$ cancel out  
and the optical theorem guarantees that the scattered state is pure, \textit{i.e.}, $\textrm{Tr}(\tilde{\rho}^2)=1$.

One of the strongest necessary and sufficient separability conditions is known as the positivity of partial transpose (PPT) criterion\cite{PhysRevLett.77.1413,HORODECKI19961}. It states that the state described by the density matrix of a two-part composite system, $\rho_{AB}$, is not-separable (entangled) if at least one eigenvalue of the partial transpose
\be
\rho^{T_B}_{AB}=(I\otimes T)(\rho_{AB})\,,
\ee
with $I$ and $T$ being, respectively,  the identity and transpose operations, 
is negative. An equivalent conditions requires that the determinant of the partial transpose of the density matrix is negative\cite{PhysRevA.77.030301},
\be\label{eq:detcon}
\textrm{Det}(\rho^{T_B}_{AB})<0\,.
\ee

A more quantitative measure of entanglement for a mixed state of two qubits (here generically indicated with $\rho$) is the concurrence\cite{Hill:1997pfa,Wootters:1997id}, given by
\be\label{eq:conc}
C(\rho)=\max\{0,\lam_1-\lam_2-\lam_3-\lam_4\}\,,
\ee
where $\lam_i$ are the square roots of the non-negative eigenvalues (ordered from largest to smallest) of the Hermitian matrix $R=\rho\bar{\rho}$, where $\bar{\rho}$ is the spin-flipped state of $\rho$, defined as
\be
\bar{\rho}=(\sigma_y\otimes\sigma_y)\rho^\ast(\sigma_y\otimes\sigma_y)\,,
\ee
and $\sigma_y$ indicates the second Pauli matrix. Note that the concurrence is a good entanglement monotone, and thus one finds $C(\rho)=0$ for a non-entangled (product) state, and $C(\rho)=1$ for a maximally entangled state.

We conclude this subsection by recalling our starting observation, that one could decide to partition the total Hilbert space $\mathcal{H}_{\textrm{tot}}$
in different ways, for example into two subsystems associated with the two scattered particles: $\mathcal{H}_{\textrm{tot}}=\mathcal{H}_{A}\,\otimes\,\mathcal{H}_{B}$, where each $\mathcal{H}_{A,B}=L^2(\mathbb{R}^3)\otimes \mathbb{C}^2$. This was the preferred choice, 
for example, in Ref.\cite{Aoude:2024xpx}. We find, however, that while the bipartite final state shows entanglement, it is not straightforward to isolate in this partition the amount of entanglement pertaining specifically to the flavor sector -- related potentially to the properties of the BSM Lagrangian -- from the entanglement between the flavor and momentum spaces, which depends on the features of the $S$-matrix. As a consequence, it becomes difficult to derive universal constraints on the couplings of the BSM model. For completeness, however, we discuss this partition in Appendix~\ref{app:redAB}.

\subsection{Final state with measured momenta\label{sec:meas_mom}}

We stress that tracing out the momentum space as was done above is consistent with the conservation of probability. The information stored in the reduced density matrix $\tilde{\rho}$ is in that case \textit{inclusive}, in the sense that the system does not know what will be the final measured momentum. However, after a measurement of final-state momenta -- say, around $\mathbf{p}_C,\mathbf{p}_D$ --
is performed in the detector, the probability 
of that particular slice of phase space becomes~1, so that a different reduced density matrix should be computed.  

The generic normalized momentum state after the measurement is
\be
| f \rangle = \left( \prod_{i=1,2}\int\frac{d^3 p_i}{(2\pi)^3}\frac{1}{\sqrt{2 E_i}}\right)
\phi_{C}(\mathbf{p}_1)\phi_D(\mathbf{p}_2)\,|\mathbf{p}_1\mathbf{p}_2\rangle
 \approx \frac{1}{\sqrt{V}}\,|\mathbf{p}_C\mathbf{p}_D\rangle\,,
\ee
for some wave packet $\phi_{C,D}$ sharply peaked about momenta $|\mathbf{p}_C\mathbf{p}_D\rangle$. In the last step we have assumed that it is possible to measure the final momenta to arbitrary precision, parameterized by the normalization $V$.  

We are now interested in projecting $|\textrm{out}\rangle$ onto $|f\rangle$. 
One finds,
\bea
|\textrm{proj}\rangle\equiv |f\rangle \langle f|\textrm{out}\rangle & = &|f\rangle \sum_{\gamma\delta} \int\int\frac{d^3 p_i}{(2\pi)^3}\frac{1}{2 E_i}\,\frac{d^3 p_j}{(2\pi)^3}\frac{1}{2 E_j}\,
\psi_{\gamma\delta}(\textbf{p}_i,\textbf{p}_j) \langle f| \textbf{p}_i\textbf{p}_j \rangle | \gamma\delta\rangle \nonumber\\
 &=&\frac{1}{V}\sum_{\gamma\delta} \psi_{\gamma\delta}(\textbf{p}_C,\textbf{p}_D)  |\textbf{p}_C\textbf{p}_D \rangle | \gamma\delta\rangle\,.
\eea
Note that this state requires normalization.

The post-measurement reduced density matrix of the flavor system can be computed straightforwardly as
\be
\tilde{\rho}_p= \frac{|\textrm{proj}\rangle\langle \textrm{proj}|}{\langle \textrm{proj}| \textrm{proj}\rangle}=\sum_{\alpha\beta,\gamma\delta}
(\tilde{\rho}_p)_{\alpha\beta,\gamma\delta}|\alpha\beta\rangle\langle\gamma\delta|\,.
\ee
In this work, we focus on the case where the measurement is made at small pseudorapidity in the detector, far away from the beamline, so that $|\mathbf{p}_C\mathbf{p}_D\rangle\neq |\mathbf{p}_A\mathbf{p}_B\rangle$.
In light of \refeq{eq:spec}, the elements of the reduced density matrix 
in the $|\alpha\beta\rangle\langle\gamma\delta|$ basis take then a simple form, 
\be\label{eq:rhoproP}
(\tilde{\rho}_p)_{\alpha\beta,\gamma\delta}=\frac{\sum_{\epsilon\rho,\tau\sigma}\mathcal{M}_{\alpha\beta,\epsilon\rho}(p_A,p_B\to p_C,p_D)\mathcal{M}^{\ast}_{\gamma\delta, \tau\sigma}(p_A,p_B\to p_C,p_D)\,a_{\epsilon\rho}\,a^{\ast}_{\tau\sigma}}{\sum_{\gamma\delta,\epsilon\rho,\tau\sigma}\mathcal{M}_{\gamma\delta,\epsilon\rho}(p_A,p_B\to p_C,p_D)\mathcal{M}^{\ast}_{\gamma\delta, \tau\sigma}(p_A,p_B\to p_C,p_D)\,a_{\epsilon\rho}\,a^{\ast}_{\tau\sigma}}\,.
\ee

The state given in \refeq{eq:rhoproP} is pure and the von Neumann entropy of the flavor-reduced density matrix, 
\be\label{eq:rhoredP}
(\tilde{\rho}_p)_{\alpha,\gamma}=\frac{\sum_{\beta,\epsilon\rho,\tau\sigma}\mathcal{M}^\ast_{\alpha\beta,\epsilon\rho}\mathcal{M}_{\gamma\beta, \tau\sigma}\,a^\ast_{\epsilon\rho}\,a_{\tau\sigma}}{\sum_{\gamma\delta,\epsilon\rho,\tau\sigma}\mathcal{M}^\ast_{\gamma\delta,\epsilon\rho}\mathcal{M}_{\gamma\delta, \tau\sigma}\,a^\ast_{\epsilon\rho}\,a_{\tau\sigma}}\,,
\ee
can be used to quantify the entanglement between the qubits of the Hilbert space $\mathcal{H}_{\textrm{flav}}$. Moreover, the eigenvalues of \refeq{eq:rhoredP} are in this case directly related to the concurrence, 
\be
\theta_{1,2}=\frac{1}{2}\left(1\pm\sqrt{1-C^2(\tilde{\rho}_p)}\right)\,,
\ee
which, for a pure bipartite state like the one in \refeq{eq:rhoproP}, is given by
\be\label{eq:con}
C(\tilde{\rho}_p)=2\,\left|\frac{\left(\sum_{\epsilon\rho}\mathcal{M}_{11,\epsilon\rho}\,a_{\epsilon\rho}\right)\,\left(\sum_{\epsilon\rho}\mathcal{M}_{22,\epsilon\rho}\,a_{\epsilon\rho}\right)-\left(\sum_{\epsilon\rho}\mathcal{M}_{12,\epsilon\rho}\,a_{\epsilon\rho}\right)\,\left(\sum_{\epsilon\rho}\mathcal{M}_{21,\epsilon\rho}\,a_{\epsilon\rho}\right)}{\sum_{\tau\sigma}|\sum_{\epsilon\rho}\mathcal{M}_{\tau\sigma,\epsilon\rho}\,a_{\epsilon\rho}|^2}\right|\,.
\ee

Note that in this case there is no information about the momentum space in the final-state density matrix and the entanglement can be quantified numerically. On the other hand, since any measurement affects the properties of a quantum state, it may not be possible to extract the most general information about the Lagrangian of the model in this case. 

\section{Flavor entanglement in scalar scattering\label{sec:2hdm}}

The formalism presented in the previous section is generic and can be applied to an arbitrary scattering event in an arbitrary model. In this section we are going to utilize it in the framework of a simple BSM scenario, reminiscent of the well-known 2HDM\cite{Lee:1973iz}. Entanglement in this model was recently considered in Ref.\cite{Carena:2023vjc}. This section extends the results of that work to include more scattering processes and applies a different methodology based on the expansion of the $S$-matrix at the 1-loop order following the formalism introduced in \refsec{sec:scat}. 

\subsection{The model\label{sec:model}}
The particle spectrum of the 2HDM consists of two scalar doublets of the SM $SU(2)_L$ group, $H_1$ and $H_2$, carrying hypercharge value $1/2$. The qubit is parameterized by the doublet's index $\alpha=1,2$, which we dub as the field's ``flavor.'' 

The gauge-invariant scalar potential of the model is given by\cite{Botella:1994cs}
\begin{multline}\label{eq:scapot0}
V(H_1,H_2)=\mu_1^2\,H_1^\dag H_1+\mu_2^2\,H_2^\dag H_2+\left(\mu_3^2\, H_1^\dag H_2+\textrm{H.c.}\right)\\
+\lambda_1\,(H_1^\dag H_1)^2+\lambda_2\,(H_2^\dag H_2)^2+\lambda_3\,(H_1^\dag H_1)(H_2^\dag H_2)+\lambda_4\,(H_1^\dag H_2)(H_2^\dag H_1)\\
+\left(\lambda_5\,(H_1^\dag H_2)^2+\lambda_6\,(H_1^\dag H_1)(H_1^\dag H_2)+\lambda_7\,(H_2^\dag H_2)(H_1^\dag H_2)+\textrm{H.c.}\right)\,,
\end{multline}
where $\mu^2_{1,2,3}\geq 0$ are the squared mass parameters, which we take to be positive, 
and $\lambda_{1,\dots,7}$ denote the dimensionless quartic couplings, which we assume to be real.  The scalar doublets can be explicitly decomposed into charged and neutral components, $H_\alpha=[\,h_\alpha^+,h_\alpha^0\,]^T$.

We are interested in the scattering of two complex 
fields, $h_\alpha\,h_\beta\to h_\gamma\,h_\delta$,
at high energy ($s\gg \mu^2_{1,2,3}$). Our main goal is to investigate whether the entanglement properties of the post-scattering two-scalar state give rise to specific functional constraints for the dimensionless couplings, some of which were shown in Ref.\cite{Carena:2023vjc} to induce the emergence of a global symmetry. 

The gauge symmetry of the scalar potential (\ref{eq:scapot0}) allows for several scattering processes. In the flavor basis $|11\rangle$, $|12\rangle$, $|21\rangle$, $|22\rangle$, the amplitude matrices $\mathcal{M}^{(n)}_{\alpha\beta,\gamma\delta}$ read, at the $n=0$ loop order (tree level),
\be\label{eq:scat00}
i\mathcal{M}^{(0)}(h^0h^0\to h^0h^0)=
-i\left(\begin{array}{cccc} 
4\lambda_1 & 2\lambda_6 & 2\lambda_6 & 4\lam_5\\ 
2\lambda_6 & \lambda_3+\lambda_4 & \lambda_3+\lambda_4 & 2\lambda_7 \\
2\lambda_6 & \lambda_3+\lambda_4 & \lambda_3+\lambda_4 & 2\lambda_7 \\
4\lam_5 & 2\lambda_7 & 2\lambda_7 & 4\lambda_2
\end{array}\right)\,,
\ee
\be\label{eq:scat0p}
i\mathcal{M}^{(0)}(h^+h^0\to h^+h^0)=-i\left(\begin{array}{cccc} 
2\lambda_1 & \lambda_6 & \lambda_6 & 2\lambda_5  \\ 
\lambda_6 & \lambda_3 & \lambda_4 & \lambda_7 \\
\lambda_6 & \lambda_4 & \lambda_3 & \lambda_7 \\
2\lambda_5 & \lambda_7 & \lambda_7 & 2\lambda_2
\end{array}\right),
\ee
\be\label{eq:scat0pt}
i\mathcal{M}^{(0)}(h^+ \tilde{h}^0\to h^+ \tilde{h}^0)=-i\left(\begin{array}{cccc} 
2\lambda_1 & \lambda_6 & \lambda_6 & \lambda_4  \\ 
\lambda_6 & \lambda_3 & 2 \lambda_5 & \lambda_7 \\
\lambda_6 & 2\lambda_5 & \lambda_3 & \lambda_7 \\
\lambda_4 & \lambda_7 & \lambda_7 & 2\lambda_2
\end{array}\right),
\ee
\be\label{eq:scatpm}
i\mathcal{M}^{(0)}(h^+h^-\to h^+h^-)=i\mathcal{M}^{(0)}(h^0 \tilde{h}^0\to h^0 \tilde{h}^0)=
-i\left(\begin{array}{cccc} 
4\lambda_1 & 2\lambda_6 & 2\lambda_6 & \lambda_3+\lambda_4  \\ 
2\lambda_6 &  \lambda_3+\lambda_4 & 4\lambda_5  & 2\lambda_7 \\
2\lambda_6 & 4\lambda_5 &   \lambda_3+\lambda_4  & 2\lambda_7 \\
\lambda_3+\lambda_4 & 2\lambda_7 & 2\lambda_7 & 4\lambda_2
\end{array}\right)\,,
\ee
\be\label{eq:scat00pp}
i\mathcal{M}^{(0)}(h^0 \tilde{h}^0\to h^+h^-)=i\mathcal{M}^{(0)}(h^+h^-\to h^0 \tilde{h}^0)=-i\left(\begin{array}{cccc} 
2\lambda_1 & \lambda_6 & \lambda_6 & \lambda_3\\ 
\lambda_6 & \lambda_4 & 2\lambda_5 & \lambda_7 \\
\lambda_6 & 2\lambda_5 & \lambda_4 & \lambda_7 \\
\lambda_3 & \lambda_7 & \lambda_7 & 2\lambda_2
\end{array}\right)\,,
\ee
where $\tilde{h}^0=h^{0\ast}$.

Unitarity of the $S$-matrix is enforced automatically at the next-to-leading order by the optical theorem.
We thus consider, for the cases given above, the 1-loop $\overline{MS}$ amplitudes in the $s$-channel, which carry the branch-cut singularity. 
For reasons of space, we only present here the corresponding $h^0 h^0\to h^0 h^0$ matrix. Each of the following entries should be multiplied by $(\log \tilde{s}-i\pi)$, where $\tilde{s}=s/\mu_{1,2,3}^2$ is the Mandelstam variable rescaled by the mass of the scattered particle: 
\be
\mathcal{M}^{(1)}_{11,11}(h^0h^0\to h^0h^0)=-\frac{4 \lam_1^2+4 \lam_5^2+2 \lam_6^2 }{8\pi^2}\,,
\ee
\begin{multline}
\mathcal{M}^{(1)}_{11,12}(h^0h^0\to h^0h^0)=\mathcal{M}^{(1)}_{11,21}(h^0h^0\to h^0h^0)=\mathcal{M}^{(1)}_{12,11}(h^0h^0\to h^0h^0)\\
=\mathcal{M}^{(1)}_{21,11}(h^0h^0\to h^0h^0)=-
\frac{8\lam_1\lam_6+4\left(\lam_3+\lam_4 \right)\lam_6+8\lam_5 \lam_7}{32\pi^2}\,,
\end{multline}
\be
\mathcal{M}^{(1)}_{11,22}(h^0h^0\to h^0h^0)=\mathcal{M}^{(1)}_{22,11}(h^0h^0\to h^0h^0)=
-\frac{2\left(\lam_1+\lam_2\right)\lam_5+\lam_6\lam_7}{4\pi^2}\,,
\ee
\begin{multline}
\mathcal{M}^{(1)}_{12,12}(h^0h^0\to h^0h^0)=\mathcal{M}^{(1)}_{12,21}(h^0h^0\to h^0h^0)=\mathcal{M}^{(1)}_{21,12}(h^0h^0\to h^0h^0)\\
=\mathcal{M}^{(1)}_{21,21}(h^0h^0\to h^0h^0)=-\frac{\left(\lam_3+\lam_4 \right)^2+2\lam_6^2+2\lam_7^2}{16\pi^2}\,,
\end{multline}
\begin{multline}
\mathcal{M}^{(1)}_{12,22}(h^0h^0\to h^0h^0)=\mathcal{M}^{(1)}_{21,22}(h^0h^0\to h^0h^0)=\mathcal{M}^{(1)}_{22,12}(h^0h^0\to h^0h^0)\\
=\mathcal{M}^{(1)}_{22,21}(h^0h^0\to h^0h^0)=-\frac{2\lam_5\lam_6+2\lam_2\lam_7+\left(\lam_3+\lam_4\right)\lam_7}{8\pi^2}\,,
\end{multline}
\be
\mathcal{M}^{(1)}_{22,22}(h^0h^0\to h^0h^0)=-\frac{2\lam_2^2+2\lam_5^2+\lam_7^2 }{4\pi^2}\,.
\ee
Equivalent 1-loop expressions apply to the scattering processes considered in Eqs.~(\ref{eq:scat0p})-(\ref{eq:scat00pp}).

\subsection{Entanglement creation\label{sec:encrea}}

Let us first assume that the incoming two-scalar state is a product state (no entanglement). 
We would like to answer the following questions: 1) Which of the couplings in \refeq{eq:scapot0} have the ability to entangle the final state?  2) What kind of entanglement, if any, is generated by the scattering process? (Momentum space/flavor space entanglement and/or two-qubit entanglement within the flavor space itself.) 

We begin with an initial state of the form~(\ref{eq:instate}), for example, 
\be\label{eq:inflav}
|\textrm{in}\rangle =\frac{1}{\sqrt{V}}\,|\mathbf{p}_A\mathbf{p}_B\rangle |11\rangle\,.
\ee
Performing the procedure described in \refsec{sec:sca_den}, we apply the $S$-matrix to construct the final state $|\textrm{out}\rangle$ and the density matrix $|\textrm{out}\rangle\langle\textrm{out}|$.

It is fairly straightforward to see that unitarity is maintained order-by-order in this construction. Let us make this point explicit for the $(11,11)$ entry of the $h^0 h^0\to h^0 h^0$ scattering amplitude. By inserting $\mathcal{M}^{(0)}+\mathcal{M}^{(1)}$ given in \refsec{sec:model} into \refeq{eq:Smat} and computing $S^{\dag}S$
one finds that 
\be\label{eq:unitar}
\frac{1}{2}\int d\Pi_2 \left(\mathcal{M}^{\dag}\mathcal{M}\right)_{11,11}=
\frac{\lam_1^2}{\pi}+\frac{\lam_5^2}{\pi}+\frac{\lam_6^2}{2\pi}+\mathcal{O}(\lam^3)= 2\,\textrm{Im}\,\mathcal{M}^{(0+1)}_{11,11}+\mathcal{O}(\lam^3)\,,
\ee
where $d\Pi_2$ indicates the usual two-particle phase-space measure, given in \refeq{eq:phsp} of Appendix~\ref{app:delta}, including the factor $1/2$ for identical particles. Similar expressions apply to all other entries of the amplitude matrix. 
Equation~(\ref{eq:unitar}) and its equivalent expressions 
automatically enforce the unitarity of the $S$-matrix at the desired order in perturbation theory. 

Given $|\textrm{out}\rangle\langle\textrm{out}|$, one can next compute with \refeq{eq:rhoredF} the elements of the momentum-reduced density matrix at the desired order. For $h^0 h^0\to h^0 h^0$ scattering they read,
\be\label{eq:rho11Ex}
\tilde{\rho}_{11,11}=1-\Delta\left(\frac{\lam_5^2}{\pi}+\frac{\lam_6^2}{2\pi}\right)\,,
\ee
\be
\tilde{\rho}_{11,12}=\tilde{\rho}_{11,21}=\tilde{\rho}_{12,11}^{\,\ast}=\tilde{\rho}_{21,11}^{\,\ast}=
\Delta\left( 2\,i\,\lam_6 +\frac{2\lam_1 \lam_6-\lam_3\lam_6-\lam_4 \lam_6-2\lam_5 \lam_7 }{8\pi} \right)\,,
\ee
\be
\tilde{\rho}_{11,22}=\tilde{\rho}_{22,11}^{\,\ast}=
\Delta\left( 4\,i\,\lam_5 +\frac{2\lam_1 \lam_5-2\lam_2\lam_5-\lam_6\lam_7 }{4\pi} \right)\,,
\ee
\be
\tilde{\rho}_{12,12}=\tilde{\rho}_{12,21}=\tilde{\rho}_{21,12}^{\,\ast}=\tilde{\rho}_{21,21}^{\,\ast}=\Delta\,\frac{\lam_6^2}{4\pi}\,,
\ee
\be
\tilde{\rho}_{12,22}=\tilde{\rho}_{21,22}=\tilde{\rho}_{22,12}^{\,\ast}=\tilde{\rho}_{22,21}^{\,\ast}=\Delta\,\frac{\lam_5 \lam_6}{2\pi}\,,
\ee
\be\label{eq:rho22Ex}
\tilde{\rho}_{22,22}=\Delta\,\frac{\lam_5^2}{\pi}\,.
\ee
Equivalent expressions apply to the reduced density matrices of the other scattering processes, and for different initial states.  Note that $\textrm{Tr}\,\tilde{\rho}=\tilde{\rho}_{11,11}+\tilde{\rho}_{12,12}+\tilde{\rho}_{21,21}+\tilde{\rho}_{22,22}=1$
at the required order, as expected from the optical theorem.  

The matrix defined by Eqs.~(\ref{eq:rho11Ex})-(\ref{eq:rho22Ex}) features two non-zero eigenvalues,
\bea
\theta_1&=&1-\Delta\left(\frac{\lam_5^2}{\pi}+\frac{\lam_6^2}{2\pi} \right)+16\,\Delta^2 \left(\lam_5^2+\frac{\lam_6^2}{2} \right)\,,\label{eq:eig1}\\
\theta_2&=&\Delta\left(\frac{\lam_5^2}{\pi}+\frac{\lam_6^2}{2\pi} \right)-16\,\Delta^2 \left(\lam_5^2+\frac{\lam_6^2}{2} \right)\,,\label{eq:eig2}
\eea
from which the von~Neumann entropy~(\ref{eq:vne})
of the reduced system can be computed. We thus find that the entanglement between the momentum and flavor degrees of freedom is generated in $h^0 h^0\to h^0 h^0$ scattering by the couplings $\lam_5$ and $\lam_6$. 

Incidentally, we note that 
\be
\textrm{Tr}\,\tilde{\rho}^2=1-\frac{\Delta}{\pi}\left(2\lam_5^2+\lam_6^2 \right)+16\,\Delta^2 \left(2\lam_5^2+\lam_6^2 \right)\,.\label{eq:trho2}
\ee
As was discussed below \refeq{eq:rhosq}
in \refsec{sec:traced}, the fact that $\Delta\leq 1/(16\pi)$ guarantees that the reduced density matrix is positive definite and that $0\leq \theta_{1,2}\leq1$.

Repeating the same analysis for the other states in the flavor basis, $|\textrm{in}\rangle_F =|12\rangle$, $|21\rangle$, $|22\rangle$, 
we can pinpoint the quartic couplings that individually create entanglement between the momentum and flavor Hilbert spaces. They are summarized in the second column of \reftable{tab:enttrace}. 

\begin{table}[t]
\begin{center}
\begin{tabular}{|P{0.18\textwidth}|P{0.3\textwidth}|P{0.3\textwidth}|}
\hline
& \multicolumn{2}{c|}{Entanglement creation}\\
\hline
$|\textrm{in}\rangle_F$ & momentum-flavor space & two-flavor space  \\
\hline
$|11\rangle$ & $\lam_3$, $\lam_4$, $\lam_5$, $\lam_6$ & $\lam_3$, $\lam_4$, $\lam_5$ \\
$|12\rangle$ & $\lam_3$, $\lam_4$, $\lam_5$, $\lam_6$, $\lam_7$ & $\lam_3$, $\lam_4$, $\lam_5$ \\
$|21\rangle$ & $\lam_3$, $\lam_4$, $\lam_5$, $\lam_6$, $\lam_7$ & $\lam_3$, $\lam_4$, $\lam_5$ \\
$|22\rangle$ & $\lam_3$, $\lam_4$, $\lam_5$, $\lam_7$ & $\lam_3$, $\lam_4$, $\lam_5$ \\
\hline
& \multicolumn{2}{c|}{Entanglement transformation}\\
 & \multicolumn{2}{c|}{flavor space $\to$ full Hilbert space}  \\
\hline
$\frac{1}{\sqrt{2}}(|11\rangle+|22\rangle)$ & \multicolumn{2}{c|}{$\lam_1-\lam_2$, $\lam_6+\lam_7$} \\
$\frac{1}{\sqrt{2}}(|11\rangle-|22\rangle)$ & \multicolumn{2}{c|}{$\lam_1-\lam_2$, $\lam_6-\lam_7$} \\
$\frac{1}{\sqrt{2}}(|12\rangle+|21\rangle)$ & \multicolumn{2}{c|}{$\lam_6+\lam_7$} \\
$\frac{1}{\sqrt{2}}(|12\rangle-|21\rangle)$ & \multicolumn{2}{c|}{none} \\
\hline
\end{tabular}
\caption{Quartic couplings of the scalar potential (\ref{eq:scapot0}) that create or transform entanglement in high-energy scalar scattering. The second column refers to the entanglement between the momentum and flavor Hilbert spaces, while the third one refers to the intrinsic entanglement between flavor degrees of freedom in the $\mathbb{C}^2\otimes\mathbb{C}^2$ flavor space. In the first column the initial flavor bipartite state is indicated. 
}
\label{tab:enttrace}
\end{center}
\end{table}

A close inspection of the results brings us to our first conclusion. If one starts with a state of the computational basis in $\mathcal{H}_{\textrm{flav}}$, the only two couplings of the scalar potential in \refeq{eq:scapot0} that do not induce, after any scattering process, entanglement between the flavor and momentum degrees of freedom are $\lam_1$ and $\lam_2$. Not surprisingly, those are the sole couplings of the Lagrangian for which the interaction is not flavored. 
In such a case the $S$-matrix acts on a subset of the Hilbert space as
\be
S\left(|\mathbf{p}_i\mathbf{p}_j\rangle|\alpha\alpha\rangle\right)=|\alpha\alpha\rangle S(\lam_\alpha)|\mathbf{p}_i\mathbf{p}_j\rangle\,, 
\ee
with $\alpha=1,2$.
This conclusion changes if we consider a generic product flavor state,
$|\textrm{in}\rangle_F =\sum_{\alpha\beta}a_{\alpha\beta}|\alpha\beta\rangle$, 
with all four~$a_{\alpha\beta}\neq 0$, as was done in 
Ref.\cite{Carena:2023vjc}. We find 
in that case that the von~Neumann entropy of the $\tilde{\rho}$ matrix depends on all seven $\lam_i$ couplings
in a highly non-trivial way. We evince that entanglement 
between the momentum and flavor degrees of freedom will always be generated by the scattering event in the most general case.  

A more interesting question is whether the scattering of two 
particles can create entanglement between two final-state flavor qubits. To answer it, recall that $\tilde{\rho}$ is in general not pure, but rather a statistical mixture of two pure eigenstates with probability $\theta_1$ and $\theta_2$ given in 
Eqs.~(\ref{eq:eig1}), (\ref{eq:eig2}).
We thus calculate the concurrence, as prescribed in \refeq{eq:conc}. The spectrum
of the Hermitian $R$ matrix is dominated by one non-zero eigenvalue
whose square root gives $C(\tilde{\rho})$ at the leading order.  
For example, given initial state~(\ref{eq:inflav}), one gets  
for $h^0 h^0\to h^0 h^0$ scattering, 
\be\label{eq:con00}
C(\tilde{\rho})=\sqrt{\frac{2\,\Delta\, \lam_5^2}{\pi}+32\, \Delta^2 \lam_5^2}
\approx \sqrt{\frac{2\Delta}{\pi}}|\lam_5|\,.
\ee
As we pointed out in \refsec{sec:traced}, at the upper physicality bound, $\Delta_{\textrm{max}}=1/(16\pi)$, the flavored scattered state becomes pure and not entangled with the momentum, as confirmed by $\textrm{Tr}\,\tilde{\rho}^2=1$ in 
\refeq{eq:trho2}. However, even in that limiting case the concurrence in \refeq{eq:con00} remains non-zero and can be simply written as
\be\label{eq:conpure}
C(\tilde{\rho})=2\sqrt{(\tilde{\rho}_{11,11}+\tilde{\rho}_{12,12})(\tilde{\rho}_{21,21}+\tilde{\rho}_{22,22})-(\tilde{\rho}_{11,21}+\tilde{\rho}_{12,22})(\tilde{\rho}_{21,11}+\tilde{\rho}_{22,12})}\,.
\ee
It is a straightforward exercise to plug Eqs.~(\ref{eq:rho11Ex})-(\ref{eq:rho22Ex}) into \refeq{eq:conpure} and verify that it coincides with \refeq{eq:con00} when $\Delta=\Delta_{\textrm{max}}$.

Repeating the same analysis for the other vectors of the flavor basis,  
$|\textrm{in}\rangle_F =|12\rangle$, $|21\rangle$, $|22\rangle$, we obtain the quartic couplings that entangle the two-qubit system. They are given in the third column of \reftable{tab:enttrace}. 

We double-check our results with the PPT criterion, specifically, 
we calculate the determinant of the partial transposed of the reduced density matrix, \refeq{eq:detcon}.
For example, for the initial state~(\ref{eq:inflav}) we get
\be
\textrm{Det}(\tilde{\rho}^{T_B})=-\Delta^3\,\frac{\lam_5^4}{\pi}\,,
\ee
which agrees at the qualitative level with the concurrence~(\ref{eq:con00}). 

Comparing the second and third columns of \reftable{tab:enttrace} one can see that, when the initial state belongs to the flavor computational basis, not all of the couplings inducing entanglement between the flavor and momentum spaces
after the scattering event also induce entanglement between the two flavor qubits. 
For example, in the specific case of $h^0 h^0\to h^0 h^0$ scattering, one can see that 
Eqs.~(\ref{eq:eig1}) and (\ref{eq:eig2}) depend on $\lam_5^2$, which sends $|11\rangle\to |22\rangle$, and on $\lam_6^2$, which sends $|11\rangle\to|12\rangle$, $|21\rangle$. Conversely,
\refeq{eq:con00} depends exclusively  on $\lam_5^2$. This pattern makes us conclude that, while any type of flavored interaction will induce some entanglement between the momentum and flavor degrees of freedom, only the couplings that contribute to the creation of maximally entangled Bell states will instead generate entanglement between the two qubits at the leading order. 

Finally, we consider again the generic product flavor state
$|\textrm{in}\rangle_F =\sum_{\alpha\beta}a_{\alpha\beta}|\alpha\beta\rangle$ 
with all four~$a_{\alpha\beta}\neq 0$. For the individual scattering process considered in Ref.\cite{Carena:2023vjc} ($h^+ h^0\to h^+ h^0$) we confirm the coupling relations found in that work. However, when all other scattering types are taken into account 
we observe that, in order to avoid the generation of any entanglement of the two flavor qubits,
the following conditions should hold: 
\be\label{eq:condall}
\lam_1=\lam_2=\lam_3=\lam_4=\lam_5=0\,,\quad \lam_6=\lam_7\,.
\ee
In light of this result, we cannot confirm the
emergence of a global symmetry from a generic requirement of entanglement suppression in this model. 

\subsection{Entanglement transformation}\label{sec:entran}

Closely related to the generation of entanglement starting from a non-entangled initial state, is the opposite situation: if we start with a maximally entangled state (for example, one of the Bell states), will the interaction be able to reduce the amount of the entanglement in the final state, and if so, to what extent.

To analyze this, let us assume that the incoming state is a maximally entangled Bell state in flavor, for example, 
\be\label{eq:inbell}
|\textrm{in}\rangle =\frac{1}{\sqrt{V}}\,|\mathbf{p}_A\mathbf{p}_B\rangle
\frac{1}{\sqrt{2}}\left(|11\rangle+|22\rangle\right)\,.
\ee
Performing an analysis along the lines of \refsec{sec:encrea} for
$h^0 h^0\to h^0 h^0$ scattering we find the eigenvalues of the momentum-reduced density matrix $\tilde{\rho}$ (at $\lam^2$ and $\Delta$ order in perturbation theory),
\bea
\theta_1&=&1-\Delta\,\frac{\left(\lam_1-\lam_2\right)^2+\left(\lam_6+\lam_7\right)^2}{4\pi}
\label{eq:eig1e}\,,\\
\theta_2&=&\Delta\,\frac{\left(\lam_1-\lam_2\right)^2+\left(\lam_6+\lam_7\right)^2}{4\pi}\,.\label{eq:eig2e}
\eea
We also find from \refeq{eq:conc} the concurrence at the leading order,
\be\label{eq:conbell}
C(\tilde{\rho})=\sqrt{1-\Delta\,\frac{\left(\lam_1-\lam_2\right)^2+\left(\lam_6+\lam_7\right)^2}{2\,\pi}}\,.
\ee

Here we observe an interesting situation. We start in \refeq{eq:inbell} with a pure state that has no entanglement between the momentum and flavor subspaces, so that the von~Neumann entropy of its momentum-reduced density matrix is trivially zero. In the flavor subspace~$\mathcal{H}_{\textrm{flav}}$, on the other hand, initial state~(\ref{eq:inbell}) is a maximally entangled Bell state with maximal initial concurrence~$C(\tilde{\rho}_{\textrm{in}})=1$. After the scattering process, some amount of entanglement ``leaves'' the flavor subspace,
as implied by the concurrence in \refeq{eq:conbell}, which is not maximal. Perhaps surprisingly, some amount of entanglement is at the same time ``injected'' into the full momentum+flavor space, as implied by the appearance of a non-trivial contribution to the von~Neumann entropy stemming from the eigenvalues~(\ref{eq:eig1e}) and~(\ref{eq:eig2e}). The amount of entanglement translating from the flavor system to the full system is parameterized by the exact same coupling combination, which we refer to as entanglement \textit{modifiers}.

By repeating the analysis with different initial states and scattering particles, one obtains qualitatively the same results: some amount of entanglement shifts from the flavor space to the full Hilbert space. The couplings that act as entanglement modifiers are summarized in the bottom panel  of \reftable{tab:enttrace} for different maximally entangled initial states. To guarantee that the initial entanglement remains always intact, one should require $\lam_1=\lam_2$ and $\lam_6=\lam_7=0$. Indeed, this provides the situation in which none of the initial Bell state is transformed in the flavor space by the action of the $S$ matrix. 

Note that one exception is provided by Bell state $|\textrm{in}\rangle_F=(|12\rangle-|21\rangle)/\sqrt{2}$, for which we observe
that the entanglement level remains maximal in flavor space for all scattering processes and independently of the couplings in the Lagrangian.  
By taking a deeper look at the form of the scattering matrices in Eqs.~(\ref{eq:scat00})-(\ref{eq:scat00pp}) -- and their 1-loop extensions -- it is straightforward to check that $|\textrm{in}\rangle_F=(|12\rangle-|21\rangle)/\sqrt{2}$ is the only Bell state for which none of the entanglement modifiers given above can appear in the density matrix at the considered perturbative order. In fact, the minus sign systematically cancels adjacent repeated entries of $\lam_6$ and $\lam_7$ against each other. Thus, we can say that this initial state constitute an entanglement invariant of the Lagrangian in flavor space.

Finally, we point out that one can identify unitary transformations generated by Lagrangian couplings $\lam_6=\lam_7$ with Local Operations and Classical Communication\cite{Bennett:1996gf} in the Hilbert space $\mathcal{H}_{\textrm{flav}}$. These types of transformations cannot generate entangled states out of separable ones (see \refeq{eq:condall}), but they can transform maximally entangled states into other less entangled ones.

\subsection{Maximization of entanglement}

We have identified in \reftable{tab:enttrace}
the quartic couplings of the scalar potential that can create or modify entanglement in scattering processes. It is important to point out, however, that the actual amount of entanglement that is generated or removed
by applying the $S$-matrix is clouded by the presence of an indeterminate normalization factor of the continuous momentum space, $\Delta$. As we have mentioned in \refsec{sec:traced} and in Appendix~\ref{app:delta}, physical consistency requires that 
$\Delta$ should be perturbatively small. This means in turn that the amount of entanglement that can be generated or removed remains small even for coupling sizes at the upper bound of perturbativity. Determining its exact size, on the other hand, would require the exact knowledge of the initial state wave packet, an issue that we have sidestepped in \refeq{eq:instate}.

The issue does not appear if we project the scattered final state on particular outgoing momenta and use the reduced post-measurement density matrix of \refeq{eq:rhoredP} to quantify the entanglement of the two flavor qubits, as was done, for example, in Refs.\cite{Cervera-Lierta:2017tdt,Fedida:2022izl} for the case of electron helicity in QED. The momentum-related normalization factors, in fact, cancel out of the projected result. Ultimately, one should be aware that even when restricting ourselves to the flavor subspace $\mathcal{H}_{\textrm{flav}}$ the information extracted using \refeq{eq:rhoredP} is of a different nature with respect to the findings of Secs.~\ref{sec:encrea}
and~\ref{sec:entran}. While we were there able to formulate universal statements relating the interactions of the Lagrangian to 
the creation or removal of entanglement through the scattering process, encoded in (the perturbative expansion of) the $S$-matrix, information obtained through the projection procedure corresponds to particular momenta and may change point-by-point in momentum space. 

\begin{table}[t]
\begin{center}
\begin{tabular}{|P{0.13\textwidth}|P{0.37\textwidth}|P{0.37\textwidth}|}
\hline
$|\textrm{in}\rangle_F$ & minimal entanglement & maximal entanglement \\
\hline
&  & $\lam_1=\lam_3=\lam_4=\lam_5=0\quad$ \\
$|11\rangle$  & $2\lam_1\lam_3=\lam_6^2$,  $\,\,\lam_5=\frac{1}{2}\lam_3=\frac{1}{2}\lam_4$  & or \\
 &  & $\lam_6=0$, $\lam_1=\lam_5=\frac{1}{2}\lam_3=\frac{1}{2}\lam_4$\\
\hline
 &  & $\lam_6=\lam_7$, $\lam_3=\lam_4=\lam_5=0$\\
$|12\rangle$, $|21\rangle$ & $\lam_6\lam_7=\lam_3^2$, $\,\,\lam_5=\frac{1}{2}\lam_3=\frac{1}{2}\lam_4$ & or \\
&  & $\lam_6=\lam_7=0$, $\lam_3=\lam_4=2\lam_5$\\
\hline
 &  & $\lam_2=\lam_3=\lam_4=\lam_5=0\quad$ \\
$|22\rangle$ & $2\lam_2\lam_3=\lam_7^2$, $\,\,\lam_5=\frac{1}{2}\lam_3=\frac{1}{2}\lam_4$ & or \\
 &  & $\lam_7=0$, $\lam_2=\lam_5=\frac{1}{2}\lam_3=\frac{1}{2}\lam_4$\\
\hline
Total & 1) $\lam_3=\lam_4=\lam_5=\lam_6=\lam_7=0$& 1) $\lam_{1}=\lam_2=\lam_3=\lam_4=\lam_5=0$,  $\lam_6=\lam_7$\\
& 2) $\lam_3^2=\lam_4^2=4\lam_5^2=\lam_6\lam_7=2\lam_1\lam_2$ & 2) $\lam_1=\lam_2=\lam_5=\frac{1}{2}\lam_3=\frac{1}{2}\lam_4$, $\lam_6=\lam_7=0$\\
\hline
\end{tabular}
\caption{Conditions on the quartic couplings of the scalar potential~(\ref{eq:scapot0}) leading to minimal or maximal entanglement in the final state projected along a chosen momentum pair as explained in \refsec{sec:meas_mom}. Conditions in the ``minimal'' boxes should be considered concurrently. Conditions in the ``maximal'' boxes are  mutually exclusive. 
In the first column the initial flavor bipartite state is indicated.
The ``total'' line collects the conditions valid for all initial states.  
}
\label{tab:entproj}
\end{center}
\end{table}

We use \refeq{eq:con} to calculate the concurrence in each of the scattering processes of Eqs.~(\ref{eq:scat00})-(\ref{eq:scat00pp})
and identify the conditions corresponding to the minimization ($C(\tilde{\rho}_p)=0$) and maximization ($C(\tilde{\rho}_p)=1$) of the final-state entanglement. 
For example, given the initial state of \refeq{eq:inflav},
we obtain for the $h^0 h^0\to h^0 h^0$ scattering 
\be
C(\tilde{\rho}_p)=2\left|\frac{4\lam_1\lam_5-\lam_6^2}{4\lam_1^2 + 4\lam_5^2 +2\lam_6^2}\right|\,.
\ee
The full set of conditions is presented in \reftable{tab:entproj} for four different, pure, flavor initial states. The ``Total'' line collects the conditions valid for all initial states.  

\section{Summary and conclusions\label{sec:sum}}

In this paper, we have investigated the entanglement properties of the final state in high-energy $2\to 2$ scalar scattering, where the scalars are characterized by an internal flavor quantum number acting like a qubit. Scattered particles are described by state vectors in the Hilbert space of the momentum and flavor degrees of freedom. Working at the 1-loop order in perturbation theory, we have constructed the final-state density matrix as a function of the scattering amplitudes connecting the initial to the final state. In this construction, the unitarity of the $S$-matrix is guaranteed at the required order by the optical theorem. 

In our analysis we explored, on the one hand, the ability of quartic interactions to generate entanglement in a scattering event; on the other, the possibility of modifying (post scattering) the entanglement of a maximally entangled state. 
In the first case we found that, for the most generic initial product state, any of the non-zero couplings of the scalar potential will induce some amount of entanglement between the momentum and flavor subspaces of the Hilbert space. Some exceptions emerge, however, for example in the case that the initial particles are prepared in the same flavor state:
entanglement is not generated by the quartic couplings of the fields of the corresponding flavor. A non-entangling theory would then consist of two separate scalar sectors. 

As for the entanglement of the two flavor qubits, we find that only the couplings $\lam_3$, $\lam_4$, and $\lam_5$ of the 2HDM scalar potential~(\ref{eq:scapot0})
can entangle the system when the initial state belongs to the computational basis, as these are the only couplings that contribute to the creation of maximally entangled Bell states. For more general initial states, however, all non-zero couplings can entangle the system, with the sole exception of $\lam_6=\lam_7\neq 0$.

In an interesting finding, we have showed that scalar scattering may change the nature of the entanglement of a maximally entangled (in flavor) initial state. More precisely, we observe the transmutation of entanglement from the flavor space of two qubits into the entanglement of the full Hilbert space of flavor and momentum. This phenomenon proceeds through the couplings $\lam_1$, $\lam_{2}$, $\lam_6$, and $\lam_7$.
Incidentally, we mention in passing that we identified the Bell state $(|12\rangle-|21\rangle)/\sqrt{2}$ as a flavor-entanglement invariant of the BSM model, as it remains maximally entangled in all the scattering events considered in this study. 

We highlighted the importance of a proper assessment of the size of indeterminate normalization factors, 
as this is the only way to guarantee that the perturbative expansion preserves the positive-definiteness of the density matrix. 
To this regard, we 
reiterate that our results rely on 
the perturbative expansion of the $S$-matrix and only 
apply to the limit of high-energy scattering, $s\gg\mu^2$. A direct consequence of the perturbative treatment is that the amount of entanglement generated or destroyed in the system is always suppressed by the smallness of both the coupling constant and the indeterminate normalization factor~$\Delta$. Moreover, the amount of entanglement cannot be exactly quantified, unless the initial wave packet is completely specified. On the other hand, non-perturbative effects in 
the scattering process could alter our results. This is an interesting but highly non-trivial matter, whose investigation we leave for future work.    

Finally, other potential future lines of investigation may include, for example, the analysis of the 
entanglement properties of three-point interactions and scalar masses in the broken symmetry phase of the scalar potential and the application of the techniques employed in this study to other realistic BSM scenarios.

\newpage
 \begin{center}
 \textbf{ACKNOWLEDGMENTS}
 \end{center}
We would like to thank Spencer Chang and Gabriel Jacobo for pointing out an error in matrices~(\ref{eq:scat0pt})-(\ref{eq:scat00pp}). 
KK is supported in part by the National Science Centre (Poland) under the research Grant No.~2017/26/E/ST2/00470. EMS is supported in part by the National Science Centre (Poland) under the research Grant No.~2020/38/E/ST2/00126. The use of the CIS computer cluster at the National Centre for Nuclear Research in Warsaw is gratefully acknowledged.
\bigskip

\appendix
\addcontentsline{toc}{section}{Appendices}


\section{The meaning of $\boldsymbol{\Delta}$\label{app:delta}}

To understand the meaning of the indeterminate normalization factor $\Delta$ it is perhaps convenient to go back to \refeq{eq:norm}. Let us neglect for simplicity any flavor indices, $\mathcal{M}_{\alpha\beta,\gamma\delta}=\mathcal{M}$. 
One can insert \refeq{eq:wave_fun} into the first line of \refeq{eq:norm} to find that the term in parentheses in the second line originates from the initial-state wave packets,
\begin{multline}
\langle\textrm{out}|\textrm{out}\rangle \supset \int\int\frac{d^3 p_i}{(2\pi)^3}\frac{1}{\sqrt{2 E_i}}\frac{d^3 p_j}{(2\pi)^3}\frac{1}{\sqrt{2 E_j}}\,\phi_A^{\ast}(\textbf{p}_i) \phi_B^{\ast}(\textbf{p}_j)\\
\times \left[\int \int \frac{d^3 p_a}{(2\pi)^3}\frac{1}{\sqrt{2 E_a}}\frac{d^3 p_b}{(2\pi)^3}\frac{1}{\sqrt{2 E_b}}\,\phi_A(\textbf{p}_a) \phi_B(\textbf{p}_b)\right.\\
\left. (2\pi)^4 \delta^4 (p_a+p_b-p_i-p_j)i\mathcal{M}(p_a,p_b\to p_i,p_j)\right]+\textrm{c.c.}
\end{multline}

Inserting then the explicit form~(\ref{eq:wavep}) of our wave packets in the above expression one finds 
\begin{multline}\label{eq:deldef}
\langle\textrm{out}|\textrm{out}\rangle \supset \int \int\frac{d^3 p_i}{(2\pi)^3}\frac{1}{\sqrt{2 E_i}}\frac{d^3 p_j}{(2\pi)^3}\frac{1}{\sqrt{2 E_j}}\,\phi_A^{\ast}(\textbf{p}_i) \phi_B^{\ast}(\textbf{p}_j)\\
\frac{(2\pi)^4\delta^4(p_A+p_B-p_i-p_j)i\mathcal{M}(p_A,p_B\to p_i,p_j)}{\sqrt{4E_A E_B}\,(2\pi)^3\delta^3(0)}+\textrm{c.c.}\\
\simeq \frac{1}{\sqrt{4E_A E_B}\,(2\pi)^3\delta^3(0)}\, \frac{(2\pi)^4\delta^4(p_A+p_B-p_A-p_B)i\mathcal{M}(p_A,p_B\to p_A,p_B)}{\sqrt{4E_A E_B}\,(2\pi)^3\delta^3(0)}+\textrm{c.c.}\\
=\frac{\delta(0)\,  i\mathcal{M}(p_A,p_B\to p_A,p_B)}{s\,(2\pi)^2\,\delta^3(0)}+\textrm{c.c.}=\Delta\,i 
\mathcal{M}(p_A,p_B\to p_A,p_B)+\textrm{c.c.}
\end{multline}

What is the meaning of the maximal value allowing the density matrix to be positive definite, $\Delta_{\textrm{max}}=1/(16 \pi)$? We rephrase the question by imposing
\be\label{eq:delmax}
\Delta_{\textrm{max}}=\frac{1}{2}\int d\Pi_2\equiv \frac{1}{2} \int \int \frac{d^3 p_i}{(2\pi)^3}\frac{1}{2 E_i}\frac{d^3 p_j}{(2\pi)^3}\frac{1}{2 E_j}\,
(2\pi)^4\delta^4(p_A+p_B-p_i-p_j)\,,
\ee
where we have multiplied the usual two-particle phase-space measure by a symmetry factor.
Inserting \refeq{eq:delmax} into \refeq{eq:deldef} we see that $\Delta=\Delta_{\textrm{max}}$ would imply that the initial wave packet maximizes to a constant over the full momentum space $\mathbb{R}^3\otimes\mathbb{R}^3$,
\be
\frac{1}{\sqrt{V}}\phi_A(\mathbf{p}_i)\phi_B(\mathbf{p}_j)\sqrt{4 E_i E_j}=\textrm{const.}
\ee
This is obviously not a physical situation.

To prove that the presence of a physical wave packet induces $\Delta<\Delta_{\textrm{max}}$ we have to recall a few notions on phase-space integration:
\bea
\int d\Pi_2&=&\int\int \frac{d^3 p_i}{(2\pi)^3}\frac{1}{2 E_i}\frac{d^3 p_j}{(2\pi)^3}\frac{1}{2 E_j}\,
(2\pi)^4\delta^4(p_A+p_B-p_i-p_j)\nonumber\\
 &=& \int\frac{d^3 p_i}{(2\pi)^6}\frac{1}{4 E_i E_j}\,
(2\pi)^4\delta(E_A+E_B-E_i-E_j)\nonumber\\
 &=& \int\int\frac{p_i^2\, d p_i\, d\Omega }{(2\pi)^2}\frac{1}{4 E_i E_j}\,
\delta(E_A+E_B-E_i-E_j)\,,\label{eq:phsp}
\eea
where in the last line we have defined, with slight abuse of notation, $p_i=|\mathbf{p}_i|$, the absolute value of one of the outgoing 3-momenta.

In presence of a wave packet, the last equation will be converted into 
\be
\int\int \frac{p_i^2\, d p_i\, d\Omega }{(2\pi)^2\,4 E_i E_j}\,\frac{1}{\sqrt{V}}\,\phi_A(\mathbf{p}_i)\phi_B(\mathbf{p}_A+\mathbf{p}_B-\mathbf{p}_i) \sqrt{4 E_i E_j}\,
\delta(E_A+E_B-E_i-E_j)\,.
\ee
Defining then $X=E_i+E_j$, and recalling that (see, \textit{e.g.},\cite{Halzen:1984mc})
\be
d p_i=\frac{dX}{p_i(X)}\frac{E_i E_j}{X}\,,
\ee
one gets
\begin{multline}
\int\int \frac{d\Omega\,dX\,p_i(X)}{(2\pi)^2\,4 X} \,\frac{1}{\sqrt{V}}\,\phi_A(\mathbf{p}_i)\phi_B(\mathbf{p}_A+\mathbf{p}_B-\mathbf{p}_i) X\,
\delta(\sqrt{s}-X)\\
=\int \frac{ p_i(\sqrt{s})\, d\Omega }{(2\pi)^2 4\sqrt{s}}\, 
\frac{\phi_A(\mathbf{p}_i)\phi_B(-\mathbf{p}_i)\sqrt{s}}{\sqrt{V}}\,,
\end{multline}
where the last equality applies to the center-of-mass frame. 

We finally recall that, at high energy, $2p_i/\sqrt{s}\simeq 1$, which implies 
\be
\int d\Pi_2=\int  \frac{p_i\, d\Omega}{(2\pi)^2\,4\sqrt{s}}\simeq \frac{1}{8\pi}\,.
\ee

Since in physical situations the initial wave packet 
is not constant over the solid angle~$d\Omega$, it follows that
\be\label{eq:stat}
\int \frac{ p_i\, d\Omega }{(2\pi)^2\,4\sqrt{s}}\, 
\frac{\phi_A(\mathbf{p}_i)\phi_B(-\mathbf{p}_i)\sqrt{s}}{\sqrt{V}}  < \frac{1}{8\pi}\,.
\ee
Equation~(\ref{eq:stat}) implies that $\Delta<\Delta_{\textrm{max}}$ so that a perturbative expansion of our results in powers of $\Delta$, besides $\lam$, is justified.

\section{Tracing out one particle}\label{app:redAB}

In this appendix we derive the density matrix reduced over one of the subsystems of a bipartite composite state. It means that we trace the final-state density matrix (\ref{eq:rho}) over one momentum and one flavor index: 
\begin{multline}\label{eq:rhoredB}
\tilde{\rho}_{\alpha,\gamma}(\mathbf{p}_a;\mathbf{p}_i)=\int \frac{d^3 p_j}{(2\pi)^3}\frac{1}{2 E_j}\, \sum_{\beta}\rho_{\alpha\beta,\gamma\beta}(\mathbf{p}_a,\mathbf{p}_j;\mathbf{p}_i,\mathbf{p}_j)\\
=\sum_{\beta}a_{\alpha\beta}\,a^{\ast}_{\gamma\beta}\, \sqrt{4\,E_{a} E_{i}}\,(2\pi)^3\,  \frac{\delta^3(\mathbf{p}_a-\mathbf{p}_A) \delta^3(\mathbf{p}_i-\mathbf{p}_A) }{\delta^3(0)}
+ \sqrt{\frac{1}{2E_B}} \frac{(2\pi)^4\delta^3(\mathbf{p}_a-\mathbf{p}_A)\delta^3(\mathbf{p}_i-\mathbf{p}_A)}{\delta^3(0)\sqrt{V}} \\
\times\sum_{\beta,\epsilon\rho} \left[-i\,\sqrt{2E_a}\,\mathcal{M}^{\ast}_{\gamma\beta,\epsilon\rho}(p_A,p_B\to p_i, p_B)\delta(E_A-E_i) \,a_{\alpha\beta}\,a^{\ast}_{\epsilon\rho}\right.\\
\left.+i\,\sqrt{2E_i}\,\mathcal{M}_{\alpha\beta,\epsilon\rho}(p_A,p_B\to p_a, p_B)\delta(E_A-E_a) \,a^{\ast}_{\gamma\beta}  \,a_{\epsilon\rho}\right]\\
+\int \frac{d^3 p_j}{(2\pi)^3}\frac{1}{2 E_j}
 (2\pi)^4 \delta^4(p_a+p_j-p_A-p_B)\,(2\pi)^4 \delta^4(p_i+p_j-p_A-p_B)  \\
\times\frac{1}{V}\sum_{\beta,\epsilon\rho,\tau\sigma} \mathcal{M}_{\alpha\beta,\epsilon\rho}(p_A,p_B\to p_a,p_j)
\mathcal{M}^{\ast}_{\gamma\beta,\tau\sigma}(p_A,p_B \to p_i,p_j) \,a_{\epsilon\rho} \,a^{\ast}_{\tau\sigma}\,.
\end{multline}
Note that the delta factors impose the momentum-diagonal form of the reduced density matrix, $|\textrm{out}\rangle\langle\textrm{out}|_{\textrm{red}}$. One writes
\begin{multline}\label{eq:denredsub}
|\textrm{out}\rangle\langle\textrm{out}|_\textrm{red}=\sum_{\alpha\gamma}\int\int
\frac{d^3 p_a}{(2\pi)^3}\frac{1}{2 E_a}\frac{d^3 p_i}{(2\pi)^3}\frac{1}{2 E_i}\, \tilde{\rho}_{\alpha,\gamma}\,(\mathbf{p}_a;\mathbf{p}_i)|\mathbf{p}_a \rangle |\alpha\rangle \langle\mathbf{p}_i | \langle \gamma|\\
=\sum_{\alpha\gamma,\beta}a_{\alpha\beta}\,a^{\ast}_{\gamma\beta}\frac{1}{2E_A\,(2\pi)^3\,\delta^3(0)}\,|\mathbf{p}_A \rangle |\alpha\rangle \langle\mathbf{p}_A | \langle \gamma|+ \frac{2\pi\,\delta(0)}{2E_A\,V}\,\sum_{\alpha\gamma,\beta,\epsilon\rho}\,|\mathbf{p}_A \rangle \langle\mathbf{p}_A | \\
\times\left[-i\,\mathcal{M}^{\ast}_{\gamma\beta,\epsilon\rho}(p_A,p_B\to p_A, p_B) \,a_{\alpha\beta}\,a^{\ast}_{\epsilon\rho}|\alpha\rangle \langle \gamma|
+i\,\mathcal{M}_{\alpha\beta,\epsilon\rho}(p_A,p_B\to p_A, p_B) \,a^{\ast}_{\gamma\beta}  \,a_{\epsilon\rho}|\gamma\rangle \langle \alpha|\right]\\
+\sum_{\alpha\gamma,\beta,\epsilon\rho,\tau\sigma}\int\frac{d^3 p_a}{(2\pi)^3}\frac{1}{2 E_a}\frac{[(2\pi)\delta(E_A+E_B-E_a-E_{A+B-a})]^2}{2E_a\,2E_{A+B-a}\,V}\,a_{\epsilon\rho} \,a^{\ast}_{\tau\sigma}\\
\times\mathcal{M}_{\alpha\beta,\epsilon\rho}(p_A,p_B\to p_a,p_A+p_B-p_a)
\mathcal{M}^{\ast}_{\gamma\beta,\tau\sigma}(p_A,p_B \to p_a,p_A+p_B-p_a)|\,\mathbf{p}_a \rangle |\alpha\rangle \langle\mathbf{p}_a | \langle \gamma|\,.
\end{multline}

Let us now consider the one-flavor limit of the theory, \textit{i.e.}, $\mathcal{M}_{\alpha\beta,\gamma\delta}=\mathcal{M}$. Equation~(\ref{eq:denredsub}) then reduces to
\begin{multline}\label{eq:redone}
|\textrm{out}\rangle\langle\textrm{out}|_\textrm{red}=\frac{1}{2E_A\,(2\pi)^3\,\delta^3(0)}\,|\mathbf{p}_A \rangle \langle\mathbf{p}_A | + \frac{2\pi\,\delta(0)}{2E_A\,V}\,|\mathbf{p}_A \rangle \langle\mathbf{p}_A |\,\left[-i\,\mathcal{M}^{\ast}+i\,\mathcal{M}\right]\\
+\frac{2\pi\,\delta(0)}{V}\int\frac{d^3 p_a}{(2\pi)^3}\frac{1}{2 E_a}\frac{(2\pi)\,\delta(E_A+E_B-E_a-E_{A+B-a})}{2E_a\,2E_{A+B-a}}\,|\mathcal{M}|^2\,|\,\mathbf{p}_a \rangle \langle\mathbf{p}_a |\,.
\end{multline}
Applying the optical theorem to the second addend of \refeq{eq:redone} we obtain
\begin{multline}
 \frac{2\pi\,\delta(0)}{2E_A\,V}\,\left[-i\,\mathcal{M}^{\ast}+i\,\mathcal{M}\right]=- \frac{2\pi\,\delta(0)}{2E_A\,V}\int\int\frac{d^3 p}{(2\pi)^3}\frac{1}{2 E_p}\frac{d^3 q}{(2\pi)^3}\frac{1}{2 E_q}(2\pi)^4\delta^4(p_A+p_B-p-q)\,|\mathcal{M}|^2\\
 =- \frac{2\pi\,\delta(0)}{V}\int\frac{d^3 p}{(2\pi)^3}\frac{1}{2 E_p}\frac{1}{2E_A\,2E_{A+B-p}}(2\pi)\,\delta(E_A+E_B-E_p-E_{A+B-p})\,|\mathcal{M}|^2\,,
\end{multline}
and thus
\begin{multline}\label{eq:redoneOT}
|\textrm{out}\rangle\langle\textrm{out}|_\textrm{red}=\frac{1}{2E_A\,(2\pi)^3\,\delta^3(0)}|\mathbf{p}_A \rangle \langle\mathbf{p}_A|\left[1-\Delta\int\frac{d^3 p_a}{(2\pi)^3}\frac{1}{2 E_a}\,\mathcal{I}_a\,|\mathcal{M}|^2\right]\\
+\Delta\int\frac{d^3 p_a}{(2\pi)^3}\frac{1}{2 E_a}\,\mathcal{I}_a\,\frac{1}{2E_a\,(2\pi)^3\,\delta^3(0)}\,|\mathcal{M}|^2\,|\mathbf{p}_a \rangle \langle\mathbf{p}_a|\,,
\end{multline}
where we introduced a short hand notation
\be
\mathcal{I}_a=\frac{(2\pi)\,\delta(E_A+E_B-E_a-E_{A+B-a})}{2E_{A+B-a}}\,.
\ee
  When $\mathcal{M}=\lam$, \refeq{eq:redoneOT} reproduces the results of Ref.\cite{Seki:2014cgq}. The von Neumann entropy at the order $\lam^2$ can thus be expressed as follows\cite{Seki:2014cgq},
\be
S_N(|\textrm{out}\rangle\langle\textrm{out}|_\textrm{red})= -\lam^2\log_2(\lam^2)\int\frac{d^3 p_a}{(2\pi)^3}\frac{1}{2 E_a}\mathcal{I}_a\,\Delta+\lam^2\int\frac{d^3 p_a}{(2\pi)^3}\frac{1}{2 E_a}\mathcal{I}_a\,\Delta\left[1-\log_2(\mathcal{I}_a\,\Delta)\right]\,.
\ee

Generalization to the two-flavor case is straightforward. For simplicity, let us consider the initial flavor state of the form $|11\rangle_F$, \textit{i.e.},~$a_{11}=1$ and $a_{12}=a_{21}=a_{22}=0$. Equation~(\ref{eq:denredsub}) then takes the form
\begin{multline}
|\textrm{out}\rangle\langle\textrm{out}|_\textrm{red}=\frac{1}{2E_A\,(2\pi)^3\,\delta^3(0)}|\mathbf{p}_A \rangle \langle\mathbf{p}_A|\left[| 1\rangle\langle 1|-i\,\Delta\sum_{\alpha\gamma}\left(\mathcal{M}^\ast_{\gamma1,11}| 1\rangle\langle \gamma|-\mathcal{M}_{\alpha1,11}| \alpha\rangle\langle 1|\right) \right]\\
+\Delta\int\frac{d^3 p_a}{(2\pi)^3}\frac{1}{2 E_a}\,\sum_{\alpha\gamma,\beta}\,\mathcal{I}_a\,\frac{1}{2E_a\,(2\pi)^3\,\delta^3(0)}\,\mathcal{M}_{\alpha\beta,11}\mathcal{M}_{\gamma\beta,11}^\ast\,|\mathbf{p}_a\rangle |\alpha \rangle \langle\mathbf{p}_a| \langle\gamma|\,.
\end{multline}
Diagonalization of the flavor submatrices can now be performed independently on momentum, leading to the following non-zero eigenvalues: 
\begin{eqnarray}
\theta_1&=&1-i\,\Delta\left(\mathcal{M}^\ast_{11,11}-\mathcal{M}_{11,11}\right)+\Delta\sum_{\rho}|\mathcal{M}_{1\rho,11}|^2+\Delta^2|\mathcal{M}_{21,11}|^2\,,\nonumber\\
\theta_2&=&\Delta\sum_{\rho}|\mathcal{M}_{2\rho,11}|^2-\Delta^2|\mathcal{M}_{21,11}|^2\,,\quad
 \theta_3=\Delta\sum_{\epsilon\rho}|\mathcal{M}^\ast_{\epsilon\rho,11}|^2\,,\quad   \theta_4=0\,.
\end{eqnarray}
Note that all the elements of the scattering amplitude $\mathcal{M}_{\epsilon\rho,11}$ that could transform the initial state $|11\rangle_F$ into an arbitrary flavor final state enter the entanglement entropy on equal footing and thus do not provide discriminating insight on the interactions of the Lagrangian.

\bibliographystyle{JHEP}
\bibliography{mybib}

\end{document}

%% file: macros.tex
\newcommand{\newc}{\newcommand*}

\long\def\begincomment#1\endcomment{%
        \begingroup\sf\baselineskip12pt#1\endgroup}

\newc{\etal}{\textrm{et al.}} 
\newc{\eg}{\textrm{e.g.}} 
\newc{\ie}{\textrm{i.e.}}
\newc{\etc}{\textrm{etc}}
\newc\vs{\textrm{vs.}}
\newc{\cl}{\rm {C.L.}}

\newc{\ev}{\ensuremath{\,\mathrm{eV}}}
\newc{\kev}{\ensuremath{\,\mathrm{keV}}}
\newc{\mev}{\ensuremath{\,\mathrm{MeV}}}
\newc{\gev}{\ensuremath{\,\mathrm{GeV}}}
\newc{\tev}{\ensuremath{\,\mathrm{TeV}}}
\newc{\MeV}{\mev} 
\newc{\TeV}{\tev}
\newc{\invpb}{\ensuremath{/\text{pb}}}
\newc{\invfb}{\ensuremath{\,\textrm{fb}^{-1}}}
\newc\nb{\ensuremath{\,\mathrm{nb}}} \newc\pb{\ensuremath{\,\mathrm{pb}}} \newc\fb{\ensuremath{\,\mathrm{fb}}}
\newc\pc{\ensuremath{\,\mathrm{pc}}}
\newc\kpc{\ensuremath{\,\mathrm{kpc}}}
\newc\mpc{\ensuremath{\,\mathrm{Mpc}}}
\newc\ps{\ensuremath{\,\mathrm{ps}}} 
\newc\cmeter{\ensuremath{\,\mathrm{cm}}} 
\newc\meter{\ensuremath{\,\mathrm{m}}} 
\newc\kmeter{\ensuremath{\,\mathrm{km}}}
\newc\second{\ensuremath{\,\mathrm{s}}}
\newc\msecond{\ensuremath{\,\mathrm{ms}}}
\newc\nsecond{\ensuremath{\,\mathrm{ns}}}
\newc\psecond{\ensuremath{\,\mathrm{ps}}}

\newc{\chisqmin}{\ensuremath{\chi^2_{\mathrm{min}}}}
\newc{\Delchisq}{\ensuremath{\Delta\chi^2}}
\newc{\chisq}{\ensuremath{\chi^2}}
\newc{\like}{\ensuremath{\mathcal{L}}}

\newc\lsim{\ensuremath{\mathrel{\rlap{\lower4pt\hbox{\hskip1pt$\sim$}}\raise1pt\hbox{$<$}}}}
\newc\gsim{\ensuremath{\mathrel{\rlap{\lower4pt\hbox{\hskip1pt$\sim$}}\raise1pt\hbox{$>$}}}}
\newc{\VEV}[1]{\ensuremath{\langle #1 \rangle}}
\newc{\dl}{\ensuremath{\stackrel{\leftarrow}{D}}}
\newc{\dr}{\ensuremath{\stackrel{\rightarrow}{D}}}

\newc{\bcenter}{\begin{center}}    \newc{\ecenter}{\end{center}}
\newc{\bfl}{\begin{flushleft}}    \newc{\efl}{\end{flushleft}}
\newc{\bfr}{\begin{flushright}}    \newc{\efr}{\end{flushright}}

\newc{\bi}{\begin{itemize}}
\newc{\ei}{\end{itemize}}
\newc{\bed}{\begin{description}}
\newc{\eed}{\end{description}}
\newc{\ben}{\begin{enumerate}}
\newc{\een}{\end{enumerate}}

\newc{\be}{\begin{equation}}
\newc{\ee}{\end{equation}}
\newc{\bea}{\begin{eqnarray}}
\newc{\eea}{\end{eqnarray}}
\newc{\bfle}{\begin{flalign}}
\newc{\efle}{\end{flalign}}
\newc{\ra}{\rightarrow}

\newc{\alphas}{\ensuremath{\alpha_s}}
\newc{\alphatwo}{\ensuremath{\alpha_2}}
\newc{\alphaone}{\ensuremath{\alpha_1}}
\newc{\alphai}[1]{\ensuremath{\alpha_{#1}}}
\newc{\alphaem}{\ensuremath{\alpha_{\mathrm{em}}}}
\newc{\alphaeff}{\ensuremath{\alpha_{\mathrm{eff}}}}
\newc{\sineff}{\ensuremath{\sin \theta_{\mathrm{eff}}}}
\newc{\sinsqeff}{\ensuremath{\sin^2 \theta_{\mathrm{eff}}}}
\newc{\dalphahad}{\ensuremath{\Delta \alpha_{\mathrm{had}}}}
\newc{\yt}{\ensuremath{h_t}} \newc{\yb}{\ensuremath{h_b}} \newc{\ytau}{\ensuremath{h_{\tau}}}
\newc\mz{\ensuremath{M_Z}} 
\newc\mw{\ensuremath{m_W}}
\newc\mZ{\mz}        \newc\mW{\mw}
\newc\mhsm{\ensuremath{ m_{H_{\mathrm{SM}}}}}
\newc{\mtop}{\ensuremath{ m_t}}               \newc{\mtpole}{\ensuremath{ M_t}}
\newc{\mbottom}{\ensuremath{ m_b}} 
\newc{\mtau}{\ensuremath{ m_{\tau}}}
\newc{\mt}{\mtpole}
\newc{\mb}{\mbottom} 
\newc{\rtwogg}{\ensuremath{R_{h_2}(\gamma\gamma)}}
\newc{\rtwozz}{\ensuremath{R_{h_2}(ZZ)}}
\newc{\ronegg}{\ensuremath{R_{h_1}(\gamma\gamma)}}
\newc{\ronezz}{\ensuremath{R_{h_1}(ZZ)}}
\newc{\rsiggg}{\ensuremath{R_{h_\textrm{sig}}(\gamma\gamma)}}
\newc{\rsigzz}{\ensuremath{R_{h_\textrm{sig}}(ZZ)}}
\newc{\llbar}{\ensuremath{\ell\bar{\ell}}}
\newc{\tauptaum}{\ensuremath{ \tau^+\tau^-}}
\newc{\qqbar}{\ensuremath{ q\bar{q}}} \newc{\ppbar}{\ensuremath{ p\bar{p}}}
\newc{\bbbar}{\ensuremath{ b\bar{b}}} \newc{\ttbar}{\ensuremath{ t\bar{t}}}
\newc{\ffbar}{\ensuremath{ f\bar{f}}} \newc{\tautaubar}{\ensuremath{ \tau\bar{\tau}}}

\newc{\mchi}{\ensuremath{m_\neutone}}
\newc{\squark}{\ensuremath{\tilde{q}}}
\newc{\slepton}{\ensuremath{\tilde{l}}}
\newc{\gluino}{\ensuremath{\tilde{g}}} 
\newc{\mgluino}{\ensuremath{{m_{\gluino}}}}
\newc{\wino}{\ensuremath{\tilde{W}}} 
\newc{\mwino}{\ensuremath{{m_{\wino}}}}
\newc{\tone}{\ensuremath{{\tilde{t}_1}}}
\newc{\bone}{\ensuremath{{\tilde{b}_1}}}
\newc{\Hone}{\ensuremath{{\tilde{H}_{1}}}}
\newc{\Htwo}{\ensuremath{{\tilde{H}_{2}}}}
\newc{\Hhtwo}{\ensuremath{{H_{2}}}}
\newc{\qli}{\ensuremath{{\tilde{Q}_{i}}}}
\newc{\uri}{\ensuremath{{\tilde{u}_{i}}}}
\newc{\dri}{\ensuremath{{\tilde{d}_{i}}}}
\newc{\lli}{\ensuremath{{\tilde{L}_{i}}}}
\newc{\eri}{\ensuremath{{\tilde{e}_{i}}}}

\newc{\sthw}{\ensuremath{ \sin\theta_W}}              \newc{\cthw}{\ensuremath{\cos\theta_W}}
\newc{\tanthw}{\ensuremath{ \tan\theta_W}}              \newc{\cotthw}{\ensuremath{\cot\theta_W}}
\newc{\ssqthw}{\ensuremath{\sin^2 \theta_W}}
\newc{\msbar}{\ensuremath{\overline{MS}}} \newc{\drbar}{\ensuremath{\overline{DR}}}
\newc{\mtmtsmmsbar}{\ensuremath{ m_t(m_t)^{\msbar}_{{\mathrm{SM}}}}}
\newc{\mtmtsmdrbar}{\ensuremath{ m_t(m_t)^{\drbar}_{{\mathrm{SM}}}}}
\newc{\mtmtmssmdrbar}{\ensuremath{ m_t(m_t)^{\drbar}_{{\mathrm{SUSY}}}}}
\newc{\mbmbmsbar}{\ensuremath{ m_b(m_b)^{\msbar} }}
\newc{\mbmbsmmsbar}{\ensuremath{ m_b(m_b)^{\msbar}_{{\mathrm{SM}}}}}
\newc{\mbmzsmmsbar}{\ensuremath{ m_b(\mz)^{\msbar}_{{\mathrm{SM}}}}}
\newc{\mbmzsmdrbar}{\ensuremath{ m_b(\mz)^{\drbar}_{{\mathrm{SM}}}}}
\newc{\mbmzmssmdrbar}{\ensuremath{ m_b(\mz)^{\drbar}_{{\mathrm{SUSY}}}}}
\newc{\mtaumzsmmsbar}{\ensuremath{ m_{\tau}(\mz)^{\msbar}_{{\mathrm{SM}}}}}
\newc{\mtaumzsmdrbar}{\ensuremath{ m_{\tau}(\mz)^{\drbar}_{{\mathrm{SM}}}}}
\newc{\mtaumzmssmdrbar}{\ensuremath{ m_{\tau}(\mz)^{\drbar}_{{\mathrm{SUSY}}}}}
\newc{\alphasmzms}{\ensuremath{\alpha_s(M_Z)^{\overline{MS}}}}
\newc{\alphaimzms}[1]{\ensuremath{\alpha_{#1}(M_Z)^{\overline{MS}}}}
\newc{\alphaemmz}{\ensuremath{\alpha_{\mathrm{em}}(M_Z)^{\overline{MS}}}}

\newc{\mzero}{\ensuremath{{m_0}}}
\newc{\mhalf}{\ensuremath{ m_{1/2}}}
\newc{\tanb}{\ensuremath{\tan\beta}}
\newc{\azero}{\ensuremath{ A_0}}
\newc{\signmu}{\ensuremath{\rm{sgn}\,\mu}}
\newc{\atau}{\ensuremath{{A_{\tau}}}}
\newc{\mueff}{\ensuremath{\mu_{\rm{eff}}}}
\newc{\lam}{\ensuremath{{\lambda}}}
\newc{\kap}{\ensuremath{{\kappa}}}
\newc{\alam}{\ensuremath{{A_{\lambda}}}}
\newc{\akap}{\ensuremath{{A_{\kappa}}}}
\newc{\hs}{\ensuremath{ H_s}}      
\newc{\mhs}{\ensuremath{ m_{H_s}}} 
\newc{\mgut}{\ensuremath{ M_{\rm GUT}}}
\newc{\gut}{\ensuremath{{\rm GUT}}}
\newc{\mplanck}{\ensuremath{ M_{\rm P}}}      \newc{\mpl}{\ensuremath{ M_{\rm Pl}}}
\newc{\msusy}{\ensuremath{ M_{\rm SUSY}}}      \newc{\ms}{\ensuremath{ M_{\rm S}}}
 \newc{\hu}{\ensuremath{ H_u}}       \newc{\hd}{\ensuremath{ H_d}}
 \newc{\mhu}{\ensuremath{ m_{H_u}}}       \newc{\mhd}{\ensuremath{ m_{H_d}}}
 \newc{\mhuew}{\ensuremath{ m^{\ast}_{H_u}}}       \newc{\mhdew}{\ensuremath{ m^{\ast}_{H_d}}}
 \newc{\mhuewsq}{\ensuremath{ m^{\ast\, 2}_{H_u}}}       \newc{\mhdewsq}{\ensuremath{ m^{\ast\, 2}_{H_d}}}
 \newc{\mhl}{\ensuremath{m_\hl}} 
 \newc{\mhone}{\ensuremath{m_{h_1}}} 
 \newc{\mhtwo}{\ensuremath{m_{h_2}}} 
 \newc{\mhi}{\ensuremath{m_{\tilde{h}}}} 
 \newc{\mul}{\ensuremath{m_{\tilde{u}_L}}} 
 \newc{\mbone}{\ensuremath{m_{\tilde{b}_1}}}  
 \newc{\mtone}{\ensuremath{m_{\tilde{t}_1}}} 
 \newc{\ma}{\ensuremath{m_A}} 
 \newc{\mH}{\ensuremath{m_H}} 
 \newc{\maone}{\ensuremath{m_{a_1}}} 
 \newc{\matwo}{\ensuremath{m_{a_2}}}
 \newc{\hone}{\ensuremath{h_1}}
 \newc{\htwo}{\ensuremath{h_2}}
 \newc{\aone}{\ensuremath{a_1}}
 \newc{\atwo}{\ensuremath{a_2}}
 \newc{\mqthree}{\ensuremath{m_{\tilde{Q}_3}^2}}
 \newc{\muthree}{\ensuremath{m_{\tilde{u}_3}^2}}
 \newc{\mqli}{\ensuremath{m_{\tilde{Q}_{i}}}}
 \newc{\muri}{\ensuremath{m_{\tilde{u}_{i}}}}
 \newc{\mdri}{\ensuremath{m_{\tilde{d}_{i}}}}
 \newc{\mlli}{\ensuremath{m_{\tilde{L}_{i}}}}
 \newc{\meri}{\ensuremath{m_{\tilde{e}_{i}}}}
 \newc{\ts}{\ensuremath{T_{SUSY}}}

\newc{\sigsip}{\ensuremath{\sigma^{\rm SI}_{p}}}	\newc{\sigsin}{\ensuremath{\sigma^{\rm SI}_{n}}}
\newc{\sigsdp}{\ensuremath{\sigma^{\rm SD}_{p}}}	\newc{\sigsdn}{\ensuremath{\sigma^{\rm SD}_{n}}}
\newc{\sigsi}{\ensuremath{\sigma^{\rm SI}}}	\newc{\sigsd}{\ensuremath{\sigma^{\rm SD}}}
\newc{\abund}{\ensuremath{ \Omega h^2}}
\newc{\omegadm}{\ensuremath{ \Omega_{{\rm DM}}}}     \newc{\abunddm}{\ensuremath{ \Omega_{{\rm DM}} h^2}} 
\newc{\omegam}{\ensuremath{ \Omega_{{\rm m}}}}       \newc{\abundm}{\ensuremath{ \Omega_{{\rm m}} h^2}}
\newc{\omegab}{\ensuremath{ \Omega_{{\rm b}}}}	\newc{\abundb}{\ensuremath{ \Omega_{{\rm b}} h^2}}
\newc{\omegatot}{\ensuremath{ \Omega_{{\rm TOT}}}}
\newc{\omegacdm}{\ensuremath{ \Omega_{{\rm CDM}}}}   \newc{\abundcdm}{\ensuremath{ \Omega_{{\rm CDM}} h^2}}
\newc{\omegalambda}{\ensuremath{ \Omega_{\Lambda}}} \newc{\abundlambda}{\ensuremath{ \Omega_{\Lambda} h^2}}
\newc{\omegarad}{\ensuremath{ \Omega_{{\rm rad}}}}  \newc{\abundrad}{\ensuremath{ \Omega_{{\rm rad}} h^2}}
\newc{\rhocrit}{\ensuremath{ \rho_{\rm crit}}}
\newc{\rhochi}{\ensuremath{ \rho_{\chi}}}
\newc{\abunchi}{\ensuremath{\Omega_\chi h^2}}
\newc{\abundlsp}{\ensuremath{\Omega_{\rm LSP}h^2}}

\newc{\amu}{\ensuremath{ a_{\mu}}}        \newc{\amususy}{\ensuremath{ a_{\mu}^{\mathrm{SUSY}}}}
\newc{\amuexpt}{\ensuremath{ a_{\mu}^{\mathrm{expt}}}}        \newc{\amusm}{\ensuremath{ a_{\mu}^{\mathrm{SM}}}}
\newc\deltaamu{\ensuremath{\Delta a_{\mu}}} \newc{\deltaamususy}{\ensuremath{\delta a_{\mu}^{\mathrm{SUSY}}}}
\newc\gmtwo{\ensuremath{ (g-2)_{\mu}}} 
\newc{\deltagmtwomususy}{\ensuremath{\delta\left(g-2\right)_{\mu}^{\mathrm{SUSY}}}}
\newc{\deltagmtwomu}{\ensuremath{\delta\left(g-2\right)_{\mu}}}
\newc\BR{\ensuremath{\rm BR}}
\newc\bsgamma{\ensuremath{ b\rightarrow s \gamma }}
\newc\bxsgamma{\ensuremath{\overline{B}\rightarrow X_{s}\gamma}}
\newc\brbsgamma{\ensuremath{\BR\left(\bsgamma\right)}}
\newc\brbxsgamma{\ensuremath{\BR\left(\bxsgamma\right)}}
\newc\bsmumu{\ensuremath{B_s\to\mu^+\mu^-}}
\newc\brbsmumu{\ensuremath{\BR\left(B_s\to\mu^+\mu^-\right)}}
\newc\bdmmumu{\ensuremath{\overline{B}_d\to\mu^+\mu^-}}
\newc\bbbarmix{\ensuremath{\overline{B}_s\mbox{-}B_s}}      
\newc\delmbs{\ensuremath{\Delta M_{B_s}}}
\newc{\butaunu}{\ensuremath{B_u \rightarrow \tau \nu}}
\newc{\brbutaunu}{\ensuremath{\BR\left(B_u \rightarrow \tau \nu\right)}}

\newcommand*{\reftable}[1]{Table~\ref{#1}}

        \newcommand*{\refeq}[1]{Eq.~(\ref{#1})}
     \newcommand*{\refsec}[1]{Sec.~\ref{#1}}

\newcommand*{\neutone}{\ensuremath{\chi^0_1}}

\let\oldcite\cite
\renewcommand*{\cite}{~\oldcite}

\newcommand*{\hl}{\ensuremath{h}}

